\documentclass[%
 reprint,
 amsmath,amssymb,
 aps,
]{revtex4-1}

\usepackage{graphicx}% Include figure files
\usepackage{dcolumn}% Align table columns on decimal point
\usepackage{bm}% bold math
\begin{document}

\preprint{APS/123-QED}

\title{Local degree blocking model for link prediction in complex networks}% Force line breaks with \\

\author{Zhen Liu}
 \email{quake.liu0625@gmail.com}
\author{Weike Dong}%
\author{Yan Fu}
\affiliation{%
 Web Sciences Center, School of Computer Science and Engineering, University of Electronic Science and Technology of China, Chengdu, 611731, China
}%

\date{\today}% It is always \today, today,
             %  but any date may be explicitly specified

\begin{abstract}
Recovering and reconstructing networks by accurately identifying missing and unreliable links is a vital task in the domain of network analysis and mining. In this article, by studying a specific local structure, namely a degree block having a node and its all immediate neighbors, we find it contains important statistical features of link formation for complex networks. We therefore propose a parameter-free local blocking (LB) predictor to quantitatively detect link formation in given networks via local link density calculations. The promising experimental results performed on six real-world networks suggest that the new index can outperform other traditional local similarity-based methods on most of tested networks. After further analyzing the scores' correlations between LB and two other methods, we find that the features of LB index are analogous to those of both PA index and short-path-based index, which empirically verify that large degree principle and short path principle simultaneously captured by the LB index are jointly driving link formation in complex networks.
\begin{description}
\item[PACS number(s)]
89.75.Hc

\end{description}
\end{abstract}

% PACS, the Physics and Astronomy
                             % Classification Scheme.
%\keywords{Suggested keywords}%Use showkeys class option if keyword
                              %display desired
\maketitle

%\tableofcontents

\section{\label{sec:level1}Introduction}

Studying link formation mechanism is of significance in understanding network growth and evolution. Conversely, uncovered link formation mechanism can also help and guide us to develop some useful link prediction methods\cite{liu2011link}. For example, common neighbors method originates from social balance mechanism\cite{dong2012link} and some link prediction methods based on machine learning are developed by homophily mechanism\cite{popescul2003statistical,al2006link,liben2007link}. To fulfill the task of link prediction, two types of information are utilized widely including the entity or node's property information and the network's topological information. Compared with the network's topological information, the entity's property information such as user's personal information in social networks or protein's functional attributes in protein-protein interaction networks may not be available for reasons such as privacy preservation and unreliable or absent prior biological knowledge. Therefore, the network's topological information is more preferable in most cases. Recent studies have revealed that some network's topological properties can be used to fit the probability of link formation. Clauset et al.\cite{clauset2008hierarchical} analyzed the hierarchical structure of the networks and proposed an HRG model to estimate the link probability in a dendrogram. Cannistraci et al.\cite{cannistraci2013link} took into account the local community and proposed an efficient paradigm called LCP to calculate the link similarity between pairs of nodes. Therefore, it has great potential for us to further explore and study the correlations between network's topological information and link formation.\\

So far, the network-structure-based link prediction methods can be mainly divided into two categories in the community of link prediction. One is using local network information to make a prediction whereas the other is using global network information to fulfill link prediction task. The link prediction methods using network's global information are commonly more accurate but very expensive in computation and therefore hard to be applied to large-sized networks. Guimer\`{a} and Sales-Pardo proposed a Stochastic Block Model (SBM)\cite{guimera2009missing} which is a typical link prediction model using global network information and is able to give very accurate link predictions on various kinds of networks. Liu et al.\cite{liu2013correlations} recently proposed a Fast Blocking probabilistic Model (FBM) based on the greedy strategy which can significantly improve the computational efficiency and has slightly better link prediction accuracy compared with the SBM. However, despite the significant reduction of implementation time in contrast to the SBM model, the network partitioning procedure used by the FBM still relies on the network's global information, which means that the time consumption of computations in massive networks would be remaining a bottleneck for the FBM. The second type of structure-based link prediction method is called local index (or proximity index) which has a big family including common neighbors, Adamic Adar\cite{adamic2003friends}, resource allocation\cite{zhou2009predicting}, etc. Such kind of methods commonly have weaker link prediction performance but lower time complexity than global predictors for their simple computational paradigms by merely using local information of the network and are more suitable to implement link prediction in massive networks. According to the opposite characteristics of the two categories of link prediction methods, we can summarize that an ultimate goal to design an excellent link predictor is to well handle the dilemma between computational efficiency and link prediction accuracy. Therefore, to study a simple yet superior link prediction method is of profound theoretical and practical interests in this domain. \\

Analyses on diverse link distributions in real-world networks are commonly able to inspire us to find underlying link formation mechanisms. For example, the study of community structure enables us to understand that links are more likely to cluster in the communities while
less likely to occur between the communities\cite{yan2012finding,clauset2005finding,newman2012communities,palla2005uncovering,radicchi2004defining}. In particular, we notice that, for a given network, each node and its immediate neighbors can
be naturally treated together as a specific local structure. In a sense, a central node (hub node) owning a larger degree will be more likely to be
connected by other nodes during the network evolution. This can be interpreted as a preferential attachment mechanism which
was firstly introduced by Barab\'{a}si and Albert for addressing their well-known BA network model\cite{barabasi1999emergence} and has drawn a lot of attention
by researchers from disparate scientific fields\cite{eisenberg2003preferential,capocci2006preferential,poncela2008complex,de2007preferential}. Therefore, we assume that such a special structure which can be called a
degree block would carry some useful information to reflect the trend of link formation. In this article, we inherit the computing framework of FBM and design a novel link prediction index by using the degree block information of the network. According to our assessments, the new index performs better than the traditional local indices with the same time complexity, therefore, it can easily fulfill the task of link prediction in massive networks. After a deeper analysis, we also find that the new index can capture two aspects of link formation in complex networks simultaneously, i.e. large degree principle and short path principle.

\section{Method}
If a network is partitioned into degree blocks, it's possible for us to analyze the statistical features of links in a given network. To quantitatively calculate the connecting probability for pairs of nodes in a degree block $x$, we introduce a simple measure called link density, which can be defined as\\

% Figure 1
\begin{figure}[htb]
\centering
\includegraphics[width=2in]{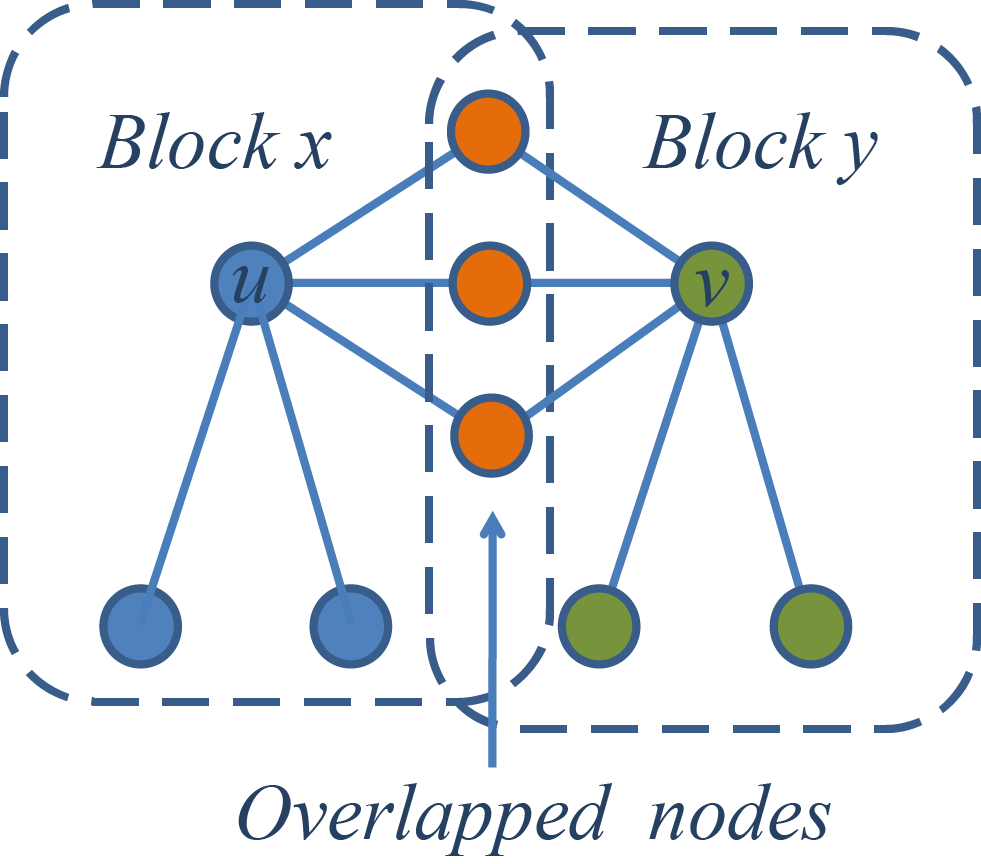}
\caption{\label{fig:figure1} Illustration of the relationship between two degree blocks. Node $u$ plus its neighbors correspond to a block $x$ and node $v$
plus its neighbors correspond to a block $y$. Note that node $u$ and node $v$ have 3 common neighbors, which
belong to both block $x$ and block $y$. In this case, block $x$ and block $y$ both have 6 members.}
\end{figure}

% Equation 1
\begin{equation}
\label{eq:eq1}
{D_x} = \frac{{\left| {{E_x}} \right|}}{{\left| {{V_x}} \right|\left| {{V_x} - 1} \right|/2}},
\end{equation}

where $|{E_x}|$ and $|{V_x}|$ are the number of links and the number of nodes in the block $x$, respectively. The denominator of Eq.
(\ref{eq:eq1}) denotes the maximal feasible number of links in the block. If the degree of central node $u$ equals $k_u$, Eq. (\ref{eq:eq1}) can be rewritten as

% Equation 2
\begin{equation}
\label{eq:eq2}
{D_x} = \frac{{k_u}}{{\left({k_u} + 1\right){k_u}/2}} =  \frac{{2}}{{{k_u} + 1}}.
\end{equation}

Moreover, for pairwise blocks, we can also calculate the link
density between two blocks $x$ and $y$, which can be defined as

% Equation 3
\begin{equation}
\label{eq:eq3}
{D_{xy}} = \frac{{\left| {{E_{xy}}} \right|}}{{\left| {{V_x}} \right|\left| {{V_y}} \right|}},
\end{equation}

where $|{E_{xy}}|$ are the number of links between block $x$ and block $y$. Likewise, the denominator of Eq. (\ref{eq:eq3}) denotes the maximal
feasible number of links between blocks $x$ and $y$. As it has a high possibility that block $x$ and block $y$ are overlapped, it is illustrated in Fig. \ref{fig:figure1} that the number of overlapped nodes will both count towards $|{V_x}|$ and $|{V_y}|$. If the degree of central node $u$ in block $x$ is $k_u$ and the degree of central node $v$ in block $y$ is $k_v$,  Eq. (\ref{eq:eq3}) can be rewritten as  

% Equation 4
\begin{equation}
\label{eq:eq4}
{D_{xy}} = \frac{{\left| {{E_{xy}}} \right|}}{{\left( {{k_u}+1} \right)\left( {{k_v}+1} \right)}}.
\end{equation}

For a network containing nodes with number $|V|$, it obviously has $|V|$ degree blocks. After the link density within and
between all blocks obtained by using Eq. (\ref{eq:eq1}) and Eq. (\ref{eq:eq3}) together in a given network, the score of link similarity for a node pair $(r,s)$ can be calculated as

% Equation 5
\begin{equation}
\label{eq:eq5}
LB(r,s) = \sum\limits_{r \in x,s \in y,x \ne y} {{D_{xy}}} +  \sum\limits_{r,s \in x} {{D_{x}}},
\end{equation}

where $x$ and $y$ denote all possible pairs of blocks which contain node $r$ and node $s$. By Eq. (\ref{eq:eq2}) and Eq. (\ref{eq:eq4}), we have that
% Equation 6
\begin{equation}
\label{eq:eq6}
\begin{split}
LB(r,s) = \sum\limits_{{
\tiny
\begin{split}
u \in N(r)-CN(r,s),\\
v \in N(s)-CN(r,s)
\end{split}
}}
{{\frac{{\left| {{E_{{b_u}{b_v}}}} \right|}}{{\left( {{k_u}+1} \right)\left( {{k_v}+1} \right)}}}}+\\
\sum\limits_{u \in CN(r,s)} {{\frac{{2}}{{{k_u} + 1}}}},
\end{split} 
\end{equation}

where $N(r)$ denotes the set of neighbors of node $r$ and $CN(r,s)$ denotes the set of common neighbors between node $r$ and node $s$. According to Eq. (\ref{eq:eq6}), the link similarity can be calculated locally by merely utilizing node $u$ and node $v$'s neighborhood information. Thus, we call this
proximity measure Local Blocking (LB) index. The whole procedure of similarity calculation for an observed network can be described in TABLE \ref{tab:table1}.\\

\begin{table*}[htb]
\caption{\label{tab:table1}Description of the algorithm of local blocking}
	\begin{ruledtabular}
		\begin{tabular}{p{0.9\linewidth}}
		\multicolumn{1}{c}{\textbf{Algorithm of local blocking}}\\
		\hline
		\textbf{Input: an observed network formalized as a $|V|\times|V|$ matrix.}\\
		\textbf{Output: link similarity matrix for all pairs of nodes in the observed network.}\\
		
		\begin{itemize}
		\item If there are $|V|$ nodes in the network, for each node $v_i$, treat it and its immediate neighbors as a degree block.
		
		\item Using Eq. (\ref{eq:eq2}) to calculate the link density for all the $|V|$ blocks. The number of operations is $|V|$.
		
		\item Using Eq. (\ref{eq:eq4}) to calculate the link density of pairwise blocks. The number of operations is $|V|(|V|-1)/2$.
		
		\item Using Eq. (\ref{eq:eq6}) to calculate the scores of link similarity for all pairs of nodes.
		\end{itemize}
		
		\end{tabular}
	\end{ruledtabular}
\end{table*}

According to the descriptions of TABLE \ref{tab:table1}, we can easily deduce that the time complexity of the algorithm is $O(|V|^2)$ which is identical to other proximity indices like CN, AA, etc.

\section{Results}
\subsection{Network data description}
In this article, six real-world networks are considered to evaluate our new link prediction index. (1) Food web\cite{christian1999organizing}: This network represents a food web in the cypress wetlands of South Florida during the wet season. (2) CN Air\cite{liu2011uncovering}: The network is extracted from the China air transportation system, which contains 121 airports and 733 airlines. (3) Infectious\cite{isella2011s}: This network describes the face-to-face infectious contacts of people during an exhibition in 2009 at a museum in Dublin. Nodes represent infected visitors; edges denote face-to-face contacts which were lasting for no less than 20 seconds. Multiple edges between two nodes are possible which denote multiple contacts between visitors. After multiple edges between each pair of nodes are incorporated to one edge, the network finally contains 410 visitors and 2396 edges. (4) C. elegans\cite{white1986structure}: A neural network of the nematode contains neurons and edges which are identified among neurons if they are connected by either a synapse or a gap junction. (5) H. friends\cite{hf2014data}: This social network contains friendships of users on the website hamsterster.com. (6) Wikivote\cite{leskovec2010predicting,leskovec2010signed}: Wikipedia is a popular online encyclopedia maintained by readers across the world.
Active users have the potential to be nominated as administrators via a voting procedure. This process can be formalized as a directed network where participating users stand for nodes and action of voting denotes a directed link. In this article, we treat it as an undirected network. The basic topology statistics of the six networks are summarized in TABLE \ref{tab:table2}.\\

\begin{table*}[htb]
\caption{\label{tab:table2}Topology statistics of six real-world networks}
\centering
\begin{ruledtabular}
    \begin{tabular}{ccccccc}
    ~ & Food web & CN Air & Infectious & C.elegans & H. friends & Wikivote \\ \hline
    $|V|$ & 128     & 121    & 410        & 297       & 1858            & 7115     \\
    $|E|$ & 2106     & 733    & 2396       & 2148      & 12533           & 103689   \\
    $D$ & 0.255  & 0.101  & 0.029      & 0.049     & 0.007           & 0.004    \\
    $C$ & 0.335  & 0.788  & 0.385      & 0.308     &  0.167          & 0.209    \\
    $\langle$$k$$\rangle$ & 32.422  & 12.116 & 11.803     & 14.465    & 13.491          & 28.324    \\
   $\langle$$d$$\rangle$ & 1.776  & 2.214  & 3.773      & 2.946     & 3.453           & 3.248

    \end{tabular}
      \footnotetext{$|V|$ is the number of vertices and $|E|$ is the number of edges in a given network. $D$ denotes the density of the network which is to calculate $2\left| E \right|/[(\left| V \right| - 1)\left| V \right|]$. A node's clustering coefficient states that, if there are $k_i$ neighbors owned by a vertex $v_i$ and the maximal feasible edges among them could be $k_i(k_i-1)/2$, the local clustering coefficient
      for the vertex $v_i$ is to calculate ${C_i} = \frac{{2\left| {{e_{jk}}:{v_j},{v_k} \in {N_i}} \right|}}{{{k_i}({k_i} - 1)}}$\cite{watts1998collective}, where $N_i$ stands for the number of neighbors of $v_i$. $C$ denotes the average clustering coefficient defined as $\frac{1}{{\left| V \right|}}\sum\nolimits_{i = 1}^{\left| V \right|} {{C_i}}$. $\langle$$k$$\rangle$ \space is the average degree of the network. Here, a shortest path is defined as a path connecting two unconnected nodes with least edges and thus the shortest path distance is the number of edges existed within the shortest path. $\langle$$d$$\rangle$ \space denotes the average shortest path distance of the network.}
\end{ruledtabular}
\end{table*}
\subsection{Metrics for evaluation}
In related literatures\cite{liu2011link,clauset2008hierarchical,guimera2009missing}, link prediction is mainly about to detect two kinds of links in networks. One kind of links are missing links in observed networks which are required to retrieve, the other kind of links are spurious links in observed networks which need to be removed. Here, we apply a widely adopted criterion called AUC (area under the receiver operating characteristic curve)\cite{hanley1983method} to evaluate the accuracy of link prediction on six networks. For missing link detection, AUC is a probability to measure whether the assigned score of a missing edge chosen at random is higher than that of a nonexistent edge chosen at random. In a test, if we ascertain that, after abundant times of comparisons independently, they include $m_1$ times that the scores of missing edges are higher than those of non-existent edges and $m_2$ times that their scores are identical, the AUC value is

% Equation 4
\begin{equation}
\label{eq:eq7}
AUC = \frac{{{m_1} + 0.5{m_2}}}{{m_1} + {m_2}}.
\end{equation}

Likewise, for spurious link prediction, AUC is a probability to measure whether a spurious edge chosen at random is assigned a lower score than a real edge chosen at random. In a test, if we ascertain that, after abundant times of randomly chosen comparisons, they include $m_1$ times that the scores of spurious edges are lower than those of real edges and $m_2$ times that their scores are identical, the AUC can be calculated by using Eq. (\ref{eq:eq7}) as well.
The set of links for a given network can be divided into a probe set and a training set. For missing link prediction, the probe set consists of some "missing" links randomly removed from the network and the training set contains the rest links of the network. For spurious link prediction, the probe set consists of some spurious links randomly added to the network and the training set consists of all the real links and added spurious links. In our tests, the fraction of links removed (added) ranges from 10\% to 90\% (The interval is 10\%). To ensure the results are of statistical significance, our tests are considered to be repeated 100 rounds, each of which corresponds to an independent division of the testing network. Each value of AUC is the average result over the 100 tests.\\

\subsection{Traditional local indices used for comparison}
Here, four traditional proximity measures are considered for performance comparison including PA, CN, AA and RA. The PA measure\cite{newman2001clustering} emphasizes that the probability of interaction between a pair of nodes is proportional to their degrees' product. Hence, it is defined as:

% Equation 5
\begin{equation}
\label{eq:eq8}
PA(r,s) = |\Gamma (r)| \cdot |\Gamma (s)|,
\end{equation}

where $|\Gamma (r)|$ is the number of node $r$'s neighbors. The CN measure states that two nodes sharing more neighbors tend to be connected which is an application of the social balance theory. The measure is formally defined as:

% Equation 6
\begin{equation}
\label{eq:eq9}
CN(r,s) = |\Gamma (r) \cap \Gamma (s)|.
\end{equation}

Different from CN method, the AA measure\cite{adamic2003friends} assumes that two nodes whose common neighbors have more neighbors tend to have lower link probability. It's formally defined as:

% Equation 7
\begin{equation}
\label{eq:eq10}
AA(r,s) = \sum\limits_{z \in \Gamma (r) \cap \Gamma (s)} {\frac{1}{{\log (|\Gamma (z)|)}}}. 
\end{equation}

The RA measure\cite{zhou2009predicting}, which is a tiny revised version of AA, is defined as:

% Equation 8
\begin{equation}
\label{eq:eq11}
RA(r,s) = \sum\limits_{z \in \Gamma (r) \cap \Gamma (s)} {\frac{1}{{|\Gamma (z)|}}}.
\end{equation}

\subsection{Results of performance comparison}
\begin{figure*}[htb]
\centering
\includegraphics[width=2.2in]{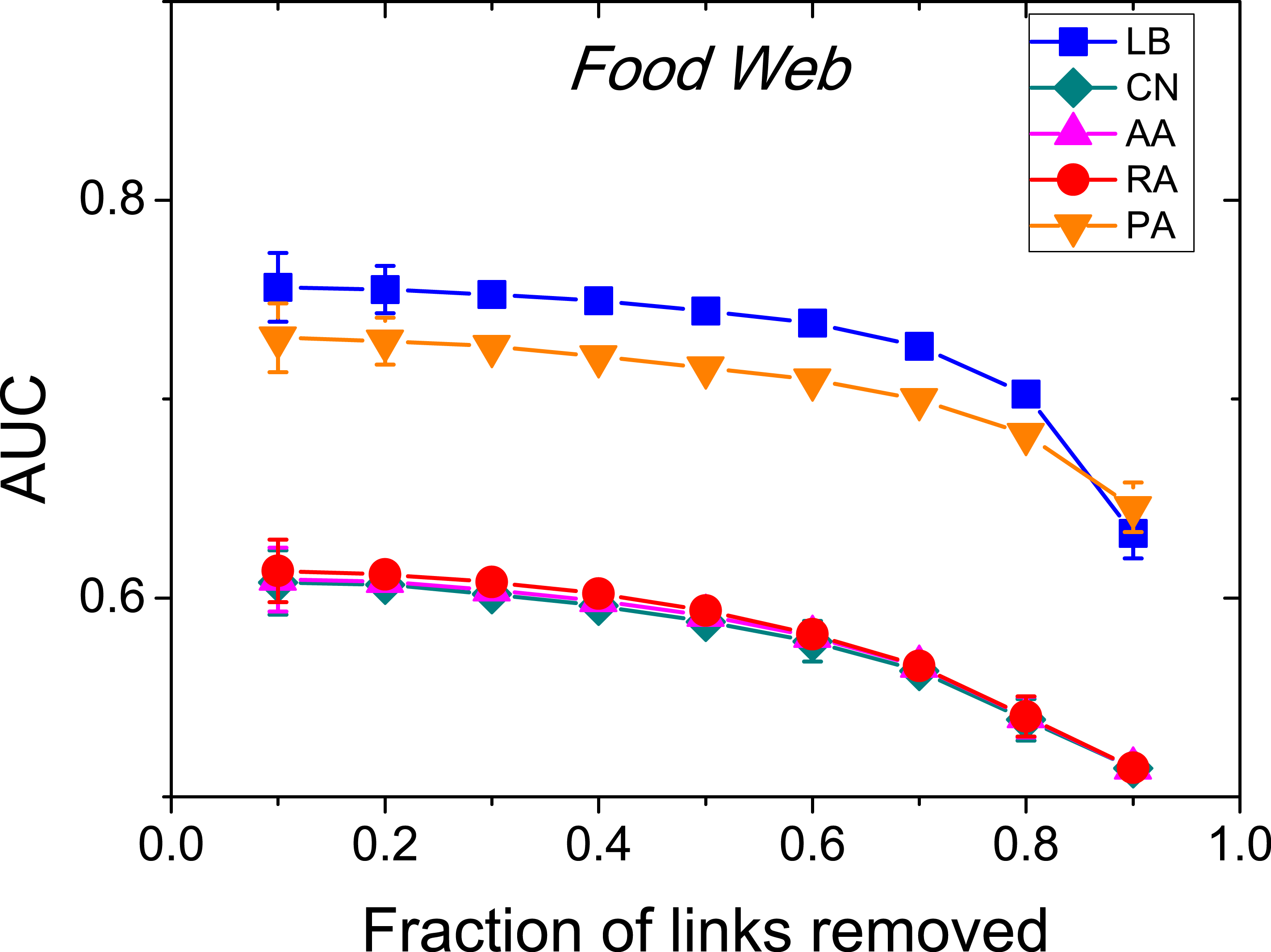}
\includegraphics[width=2.2in]{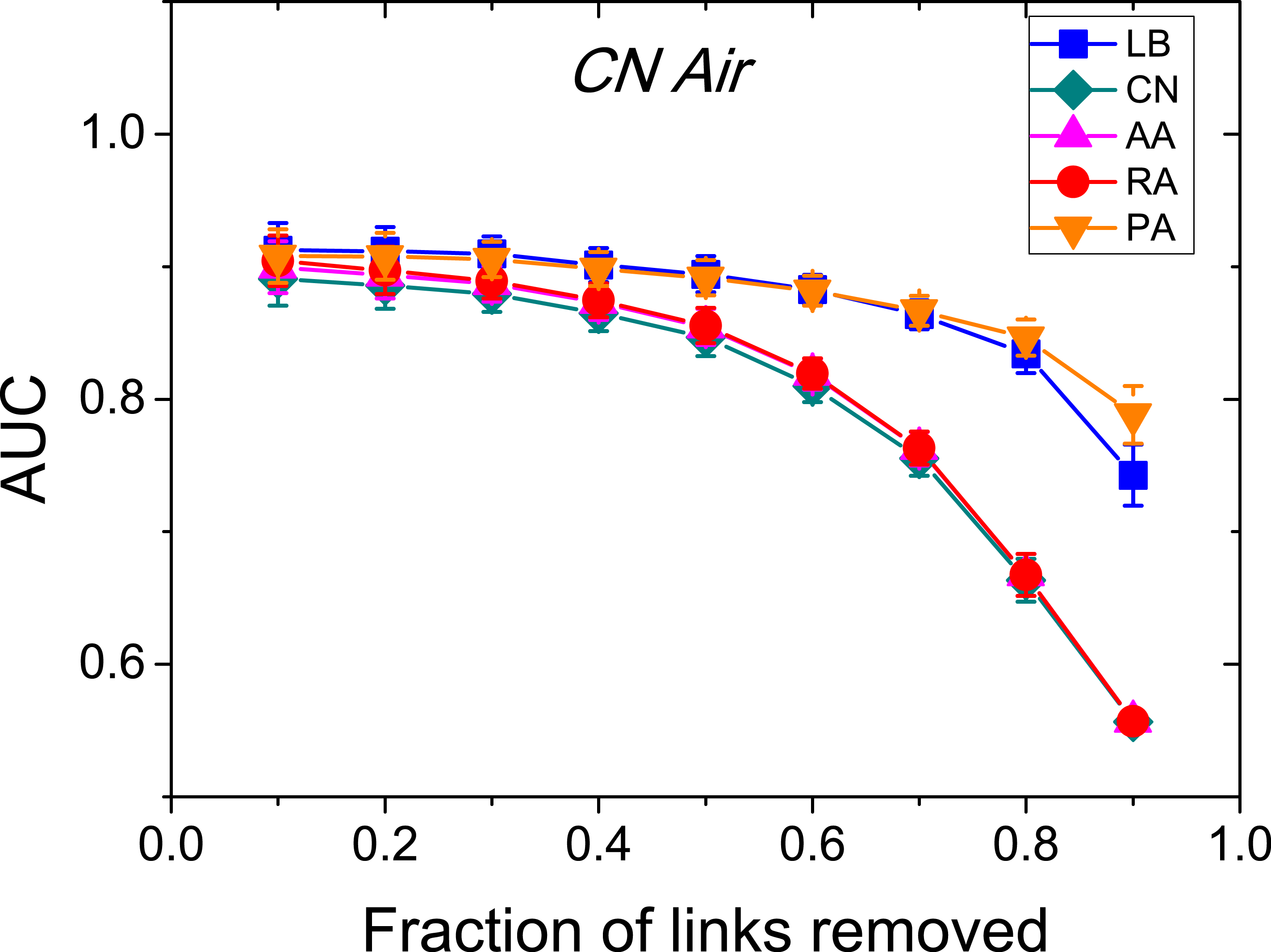}
\includegraphics[width=2.2in]{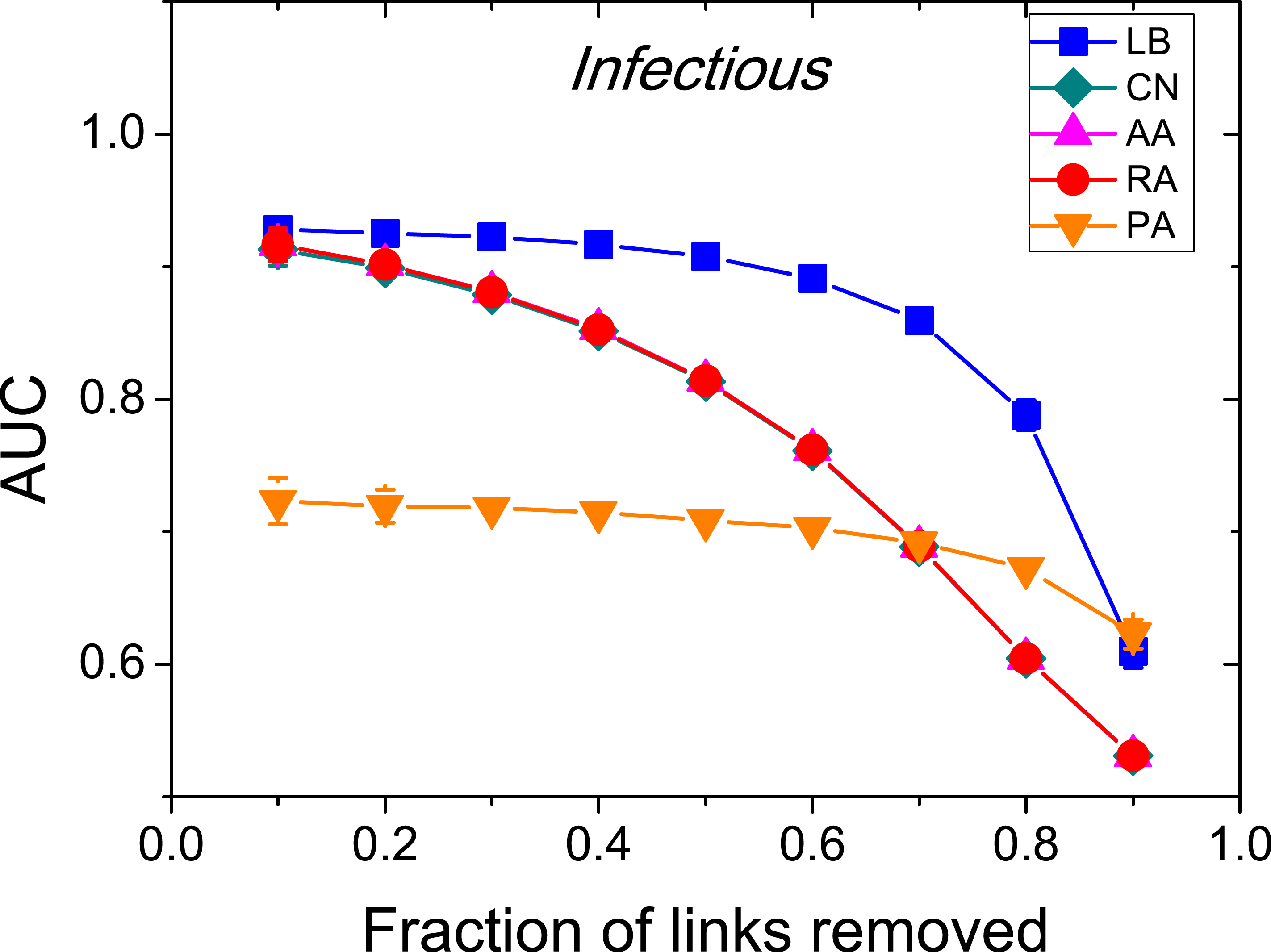}
\includegraphics[width=2.2in]{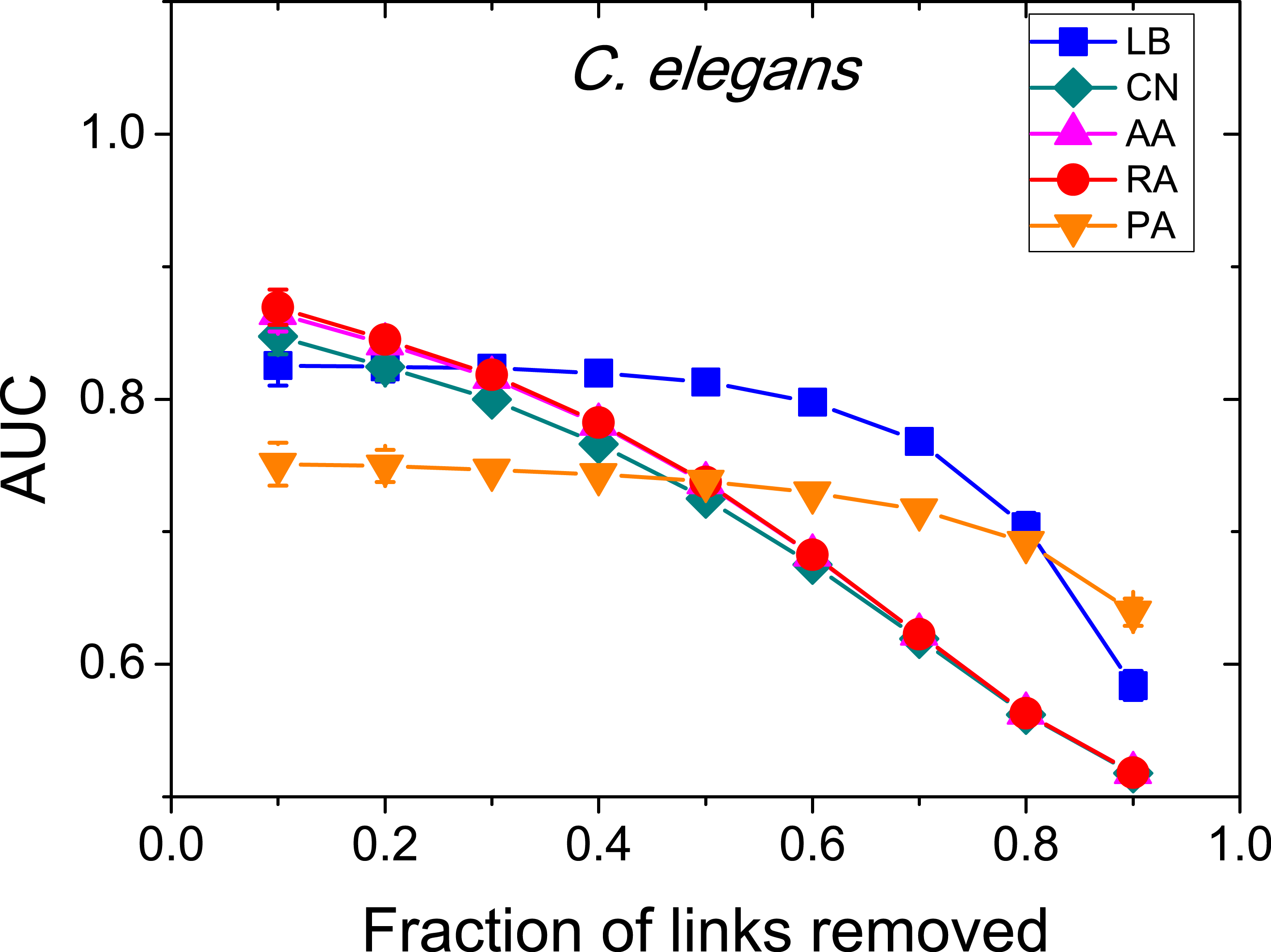}
\includegraphics[width=2.2in]{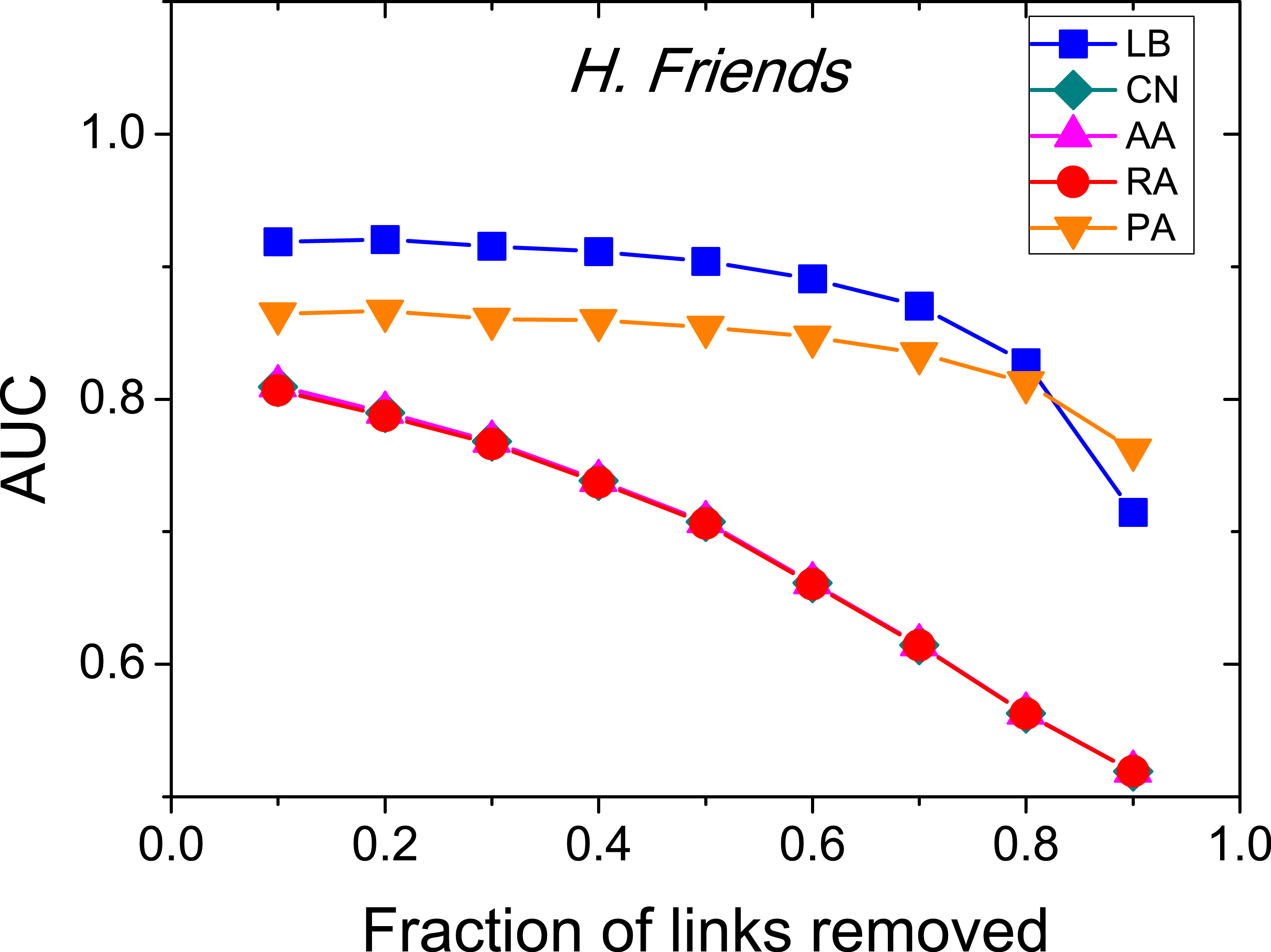}
\includegraphics[width=2.2in]{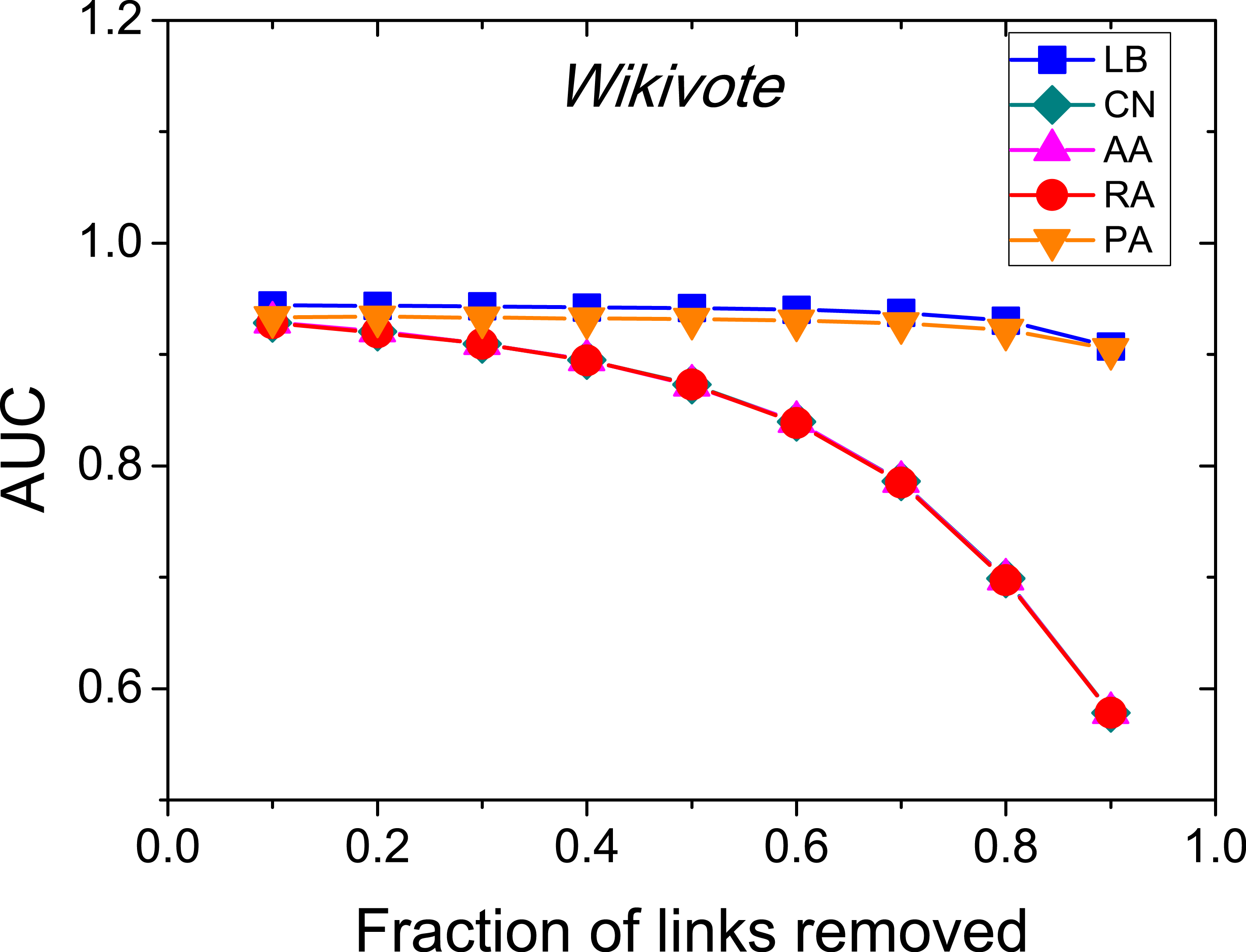}

\caption{\label{fig:figure2} AUC curves for missing link prediction. LB and four local indices are compared by the experimental results performed on six networks. Each value of AUC is a result averaged over 100 tests and the error bar corresponds to the standard deviation.}
\end{figure*}

\begin{figure*}[htb]
\includegraphics[width=2.2in]{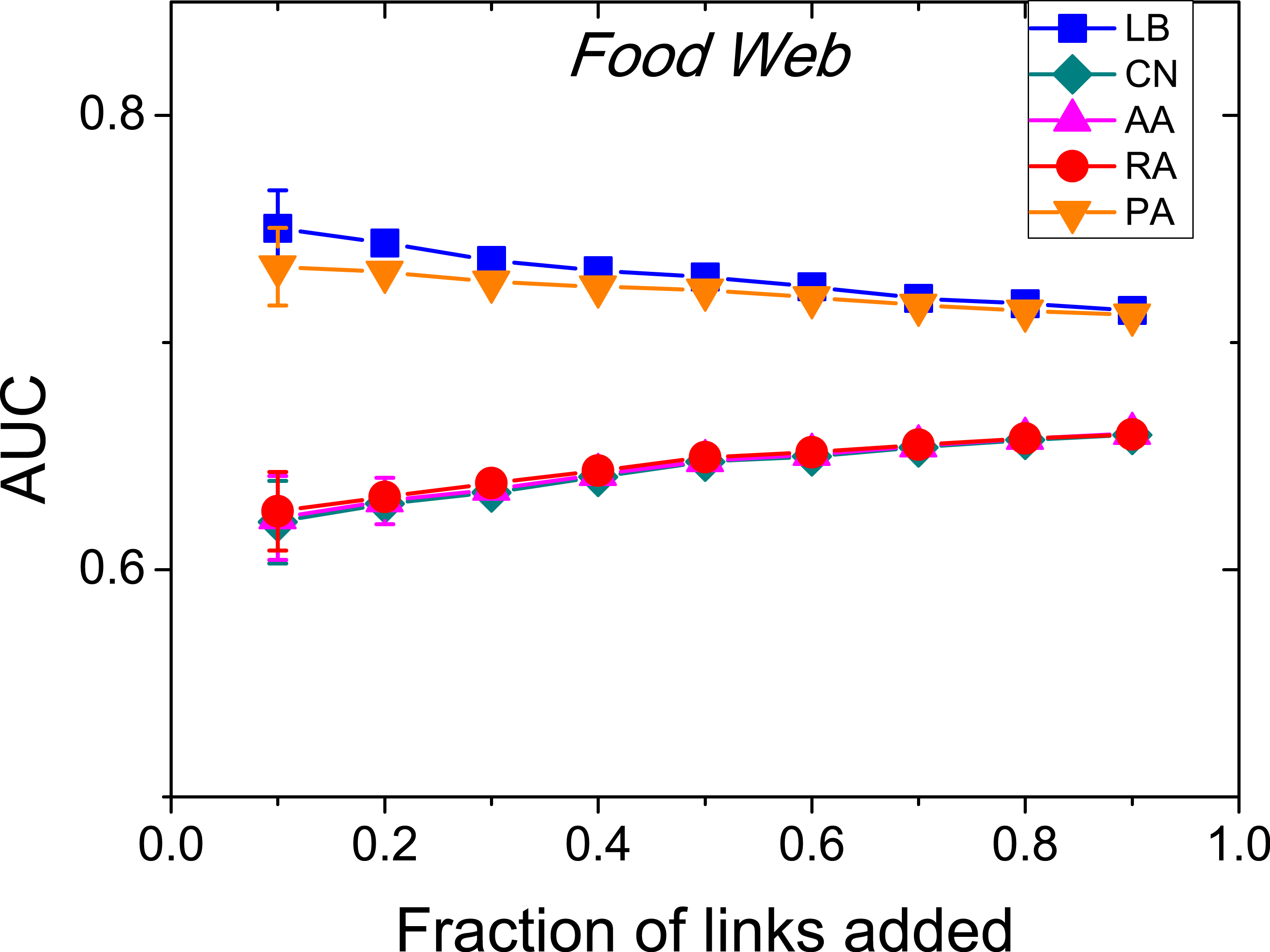}
\includegraphics[width=2.2in]{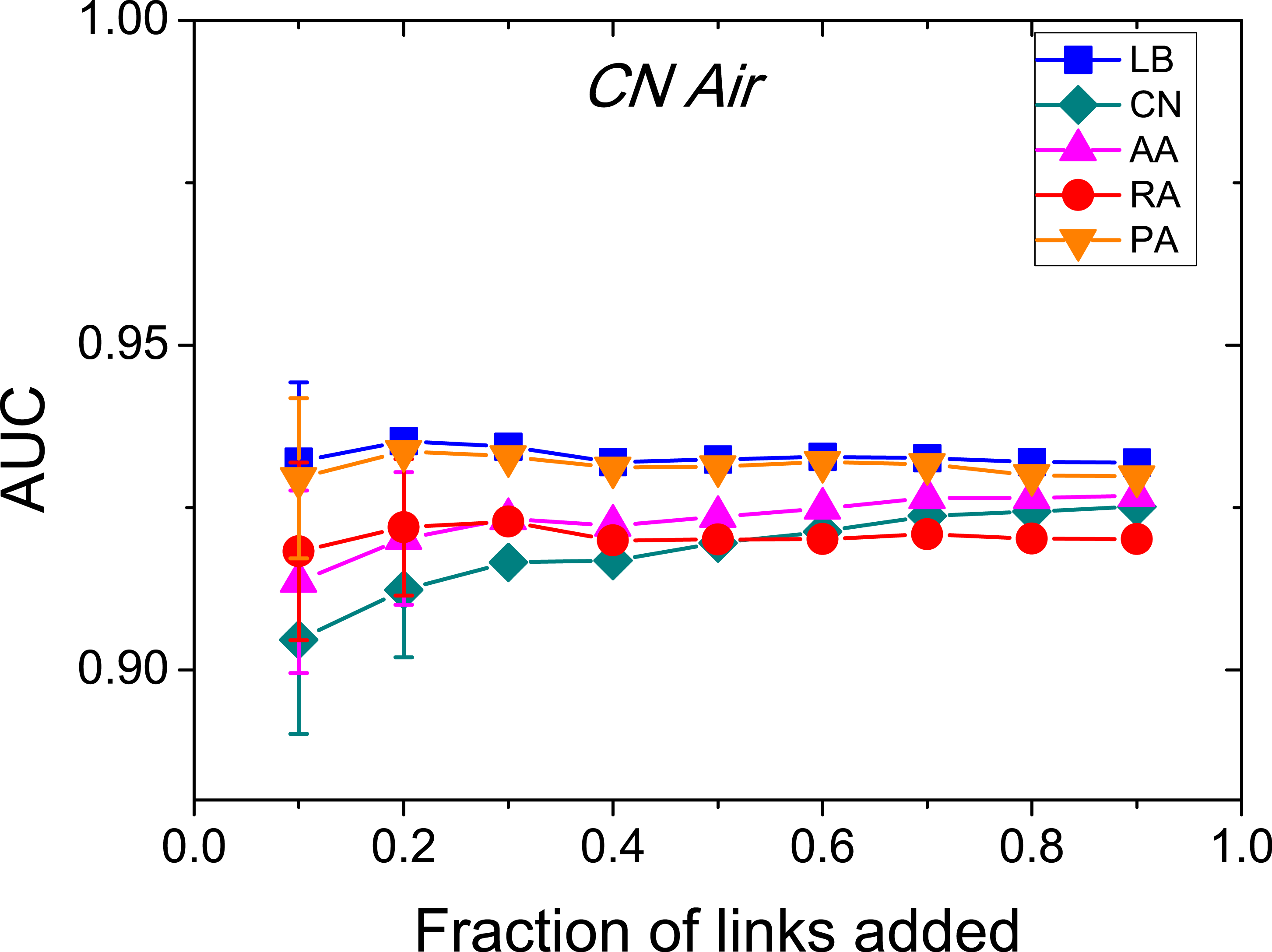}
\includegraphics[width=2.2in]{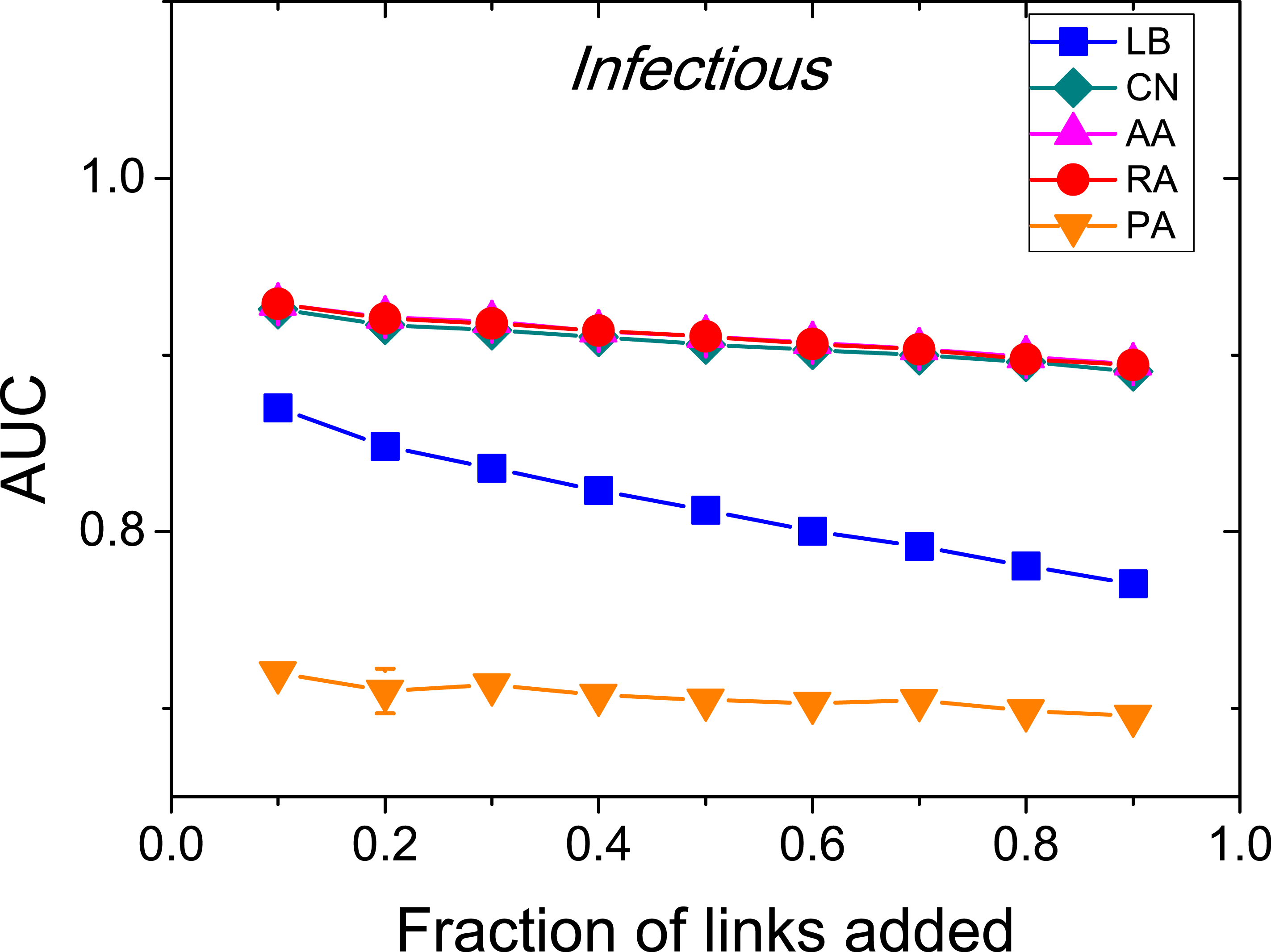}
\includegraphics[width=2.2in]{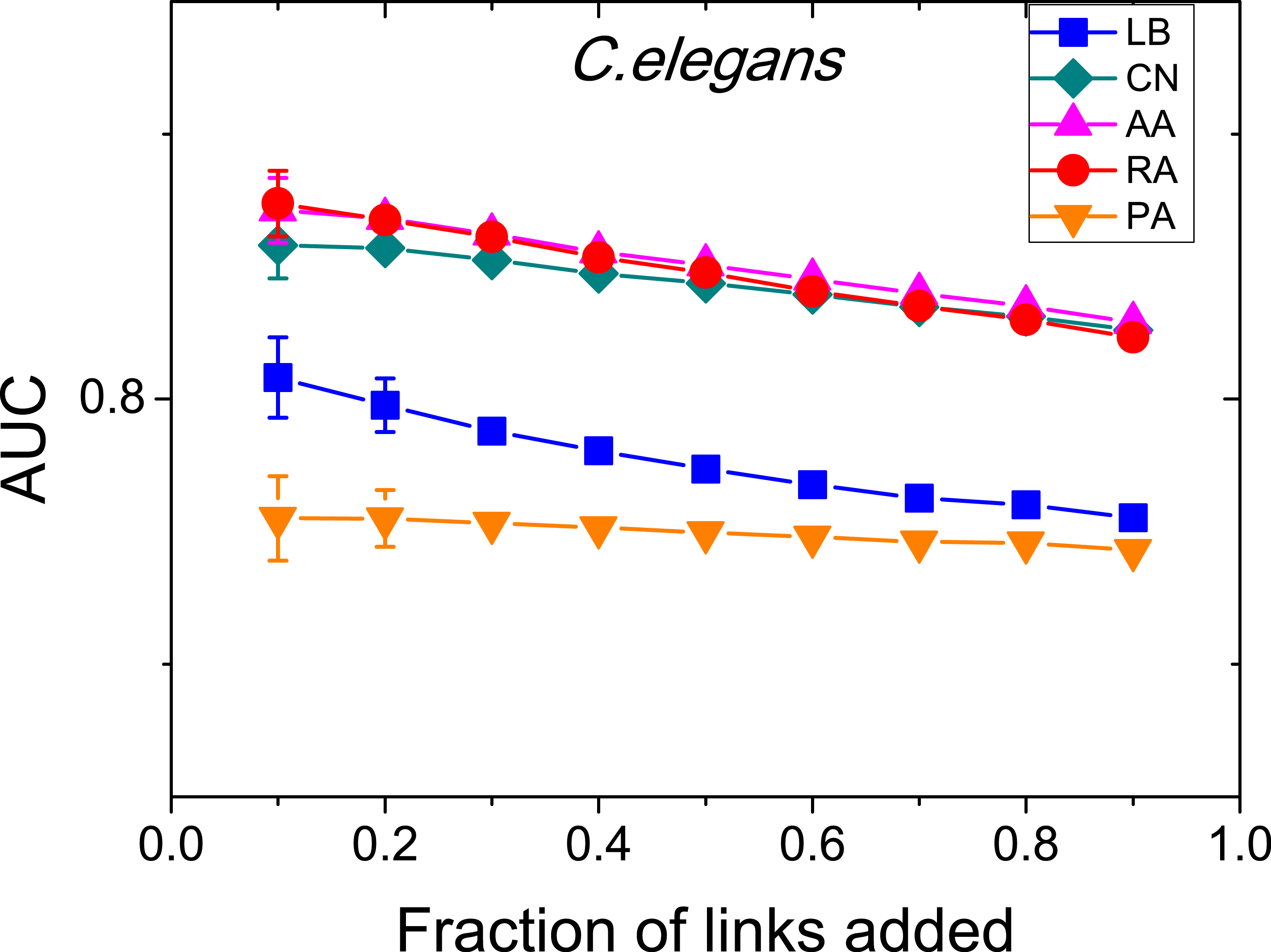}
\includegraphics[width=2.2in]{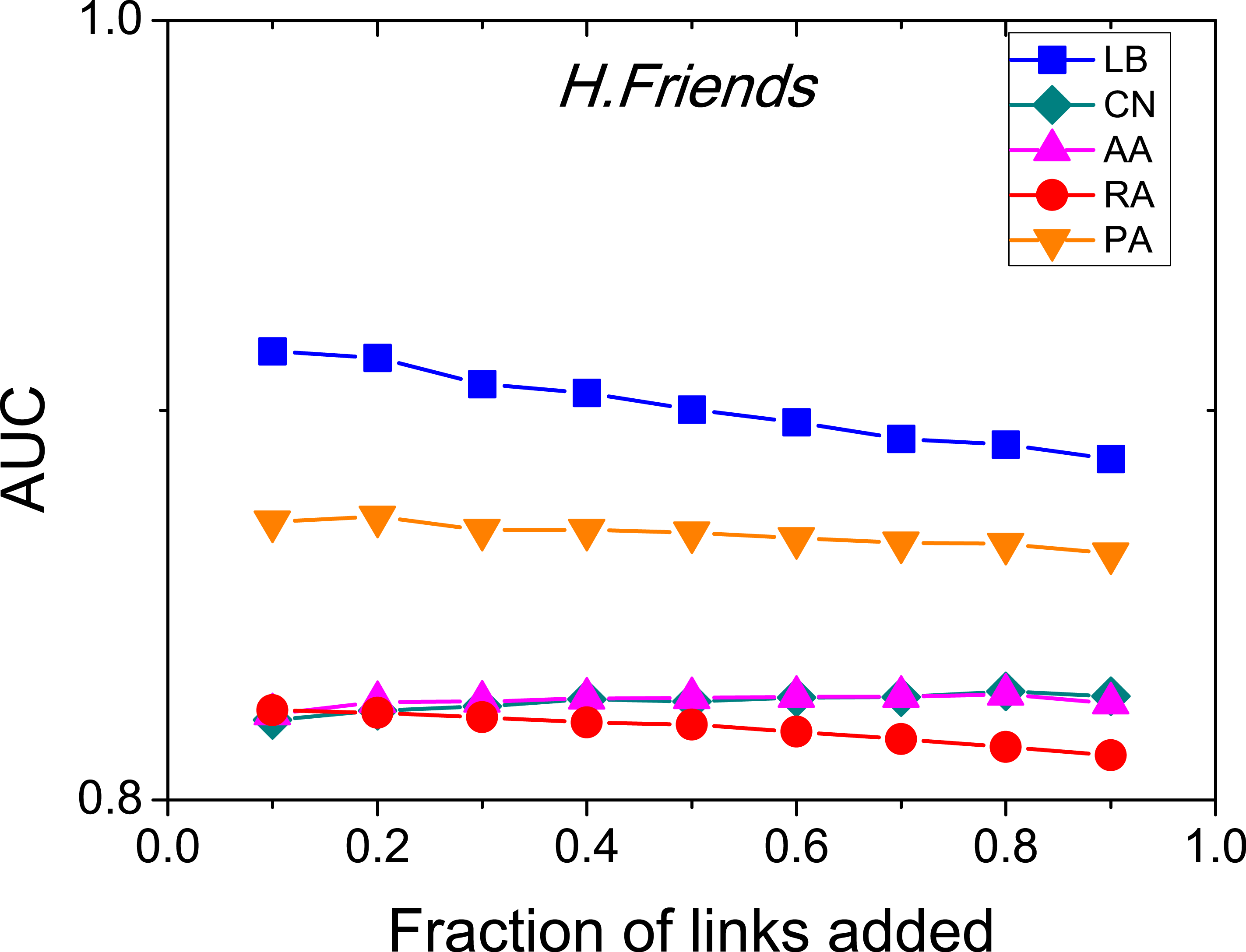}
\includegraphics[width=2.2in]{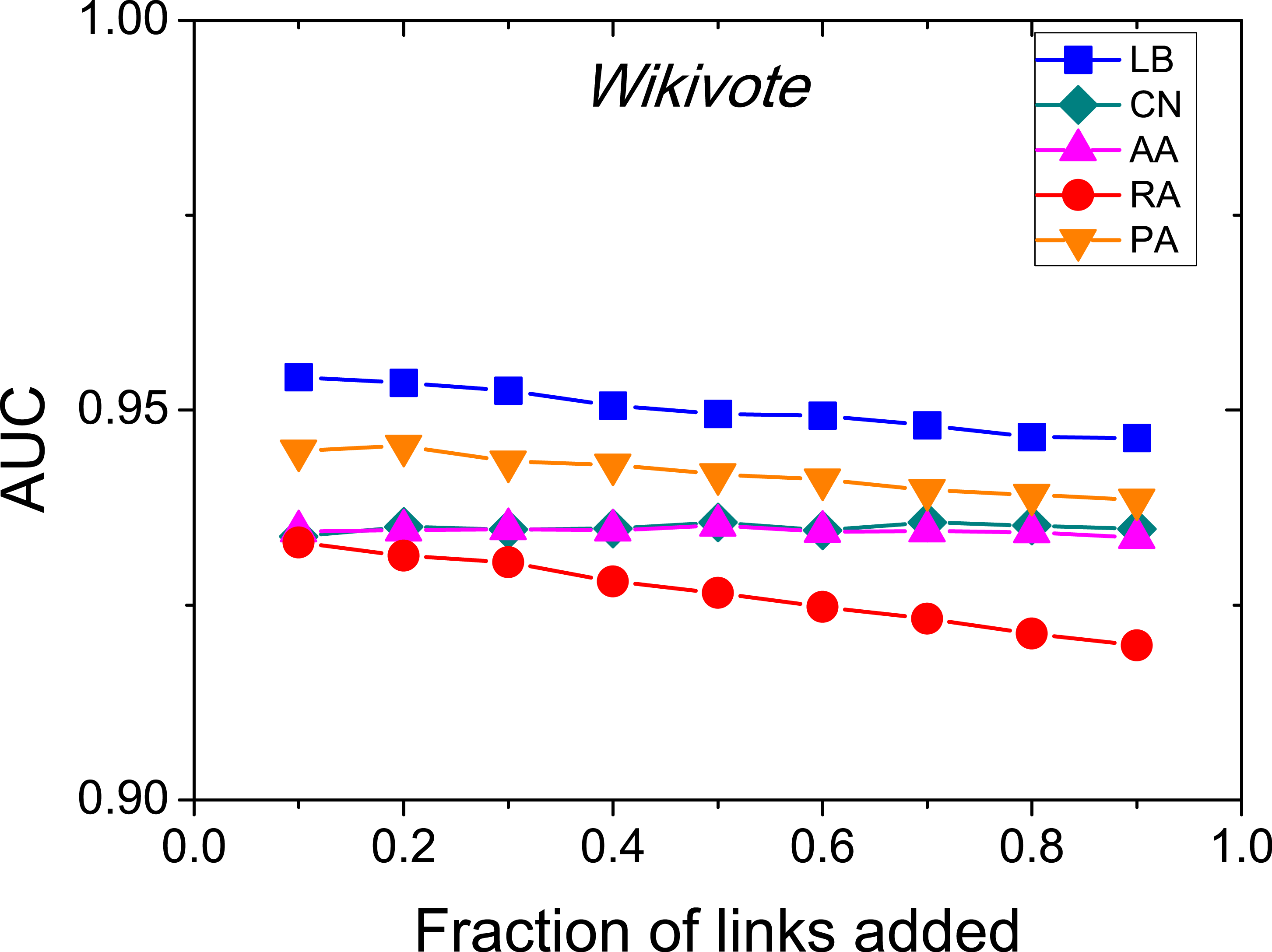}

\caption{\label{fig:figure3} AUC curves for spurious link prediction. LB and four local indices are compared by the experimental results performed on six networks. Each value of AUC is a result averaged over 100 tests and the error bar corresponds to the standard deviation.}
\end{figure*}

The accuracy comparison results are plotted in Fig. \ref{fig:figure2} and Fig. \ref{fig:figure3}. The results of missing link prediction indicate that, except for the comparable accuracy curves given by PA index in networks of CN Air and Wikivote, LB index performs better than other traditional proximity indices and significantly better than CN, AA and RA when the fraction of links removed exceeds 50\%. This suggests that LB index is pretty robust for identifying missing links with less network information and suitable for the task of link prediction on sparse networks or networks with large fraction of missing links.
For spurious link prediction, LB index is still competitive with other methods in that it performs best on four networks including Food web, CN Air, H. Friends and Wikivote. The AUC curves of missing and spurious link prediction also show that the indices of CN, AA and RA can be regarded as one type of index for their AUC curves are nearly the same on most networks.\\

\begin{figure*}[htb]
\includegraphics[width=2.2in]{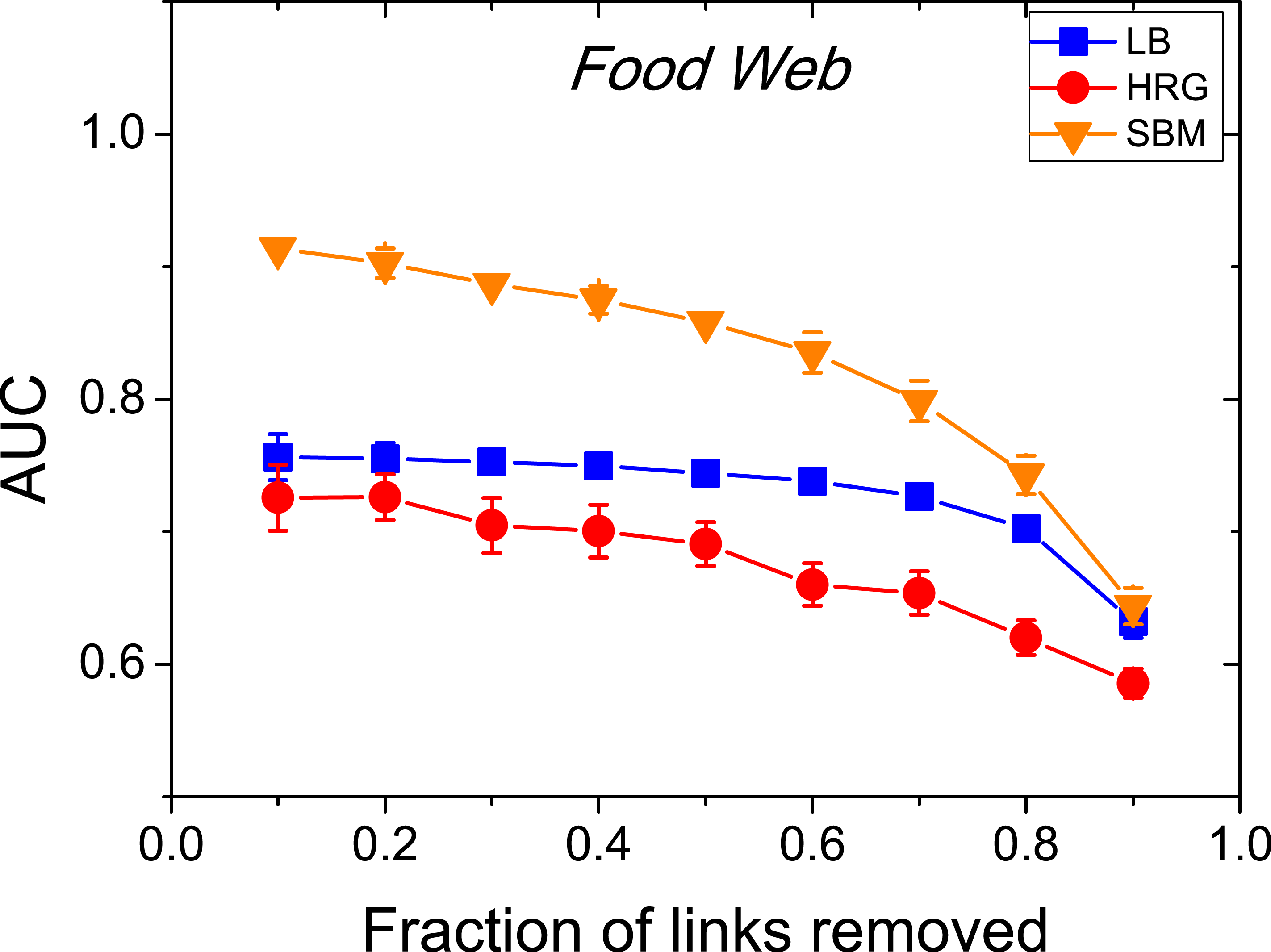}
\includegraphics[width=2.2in]{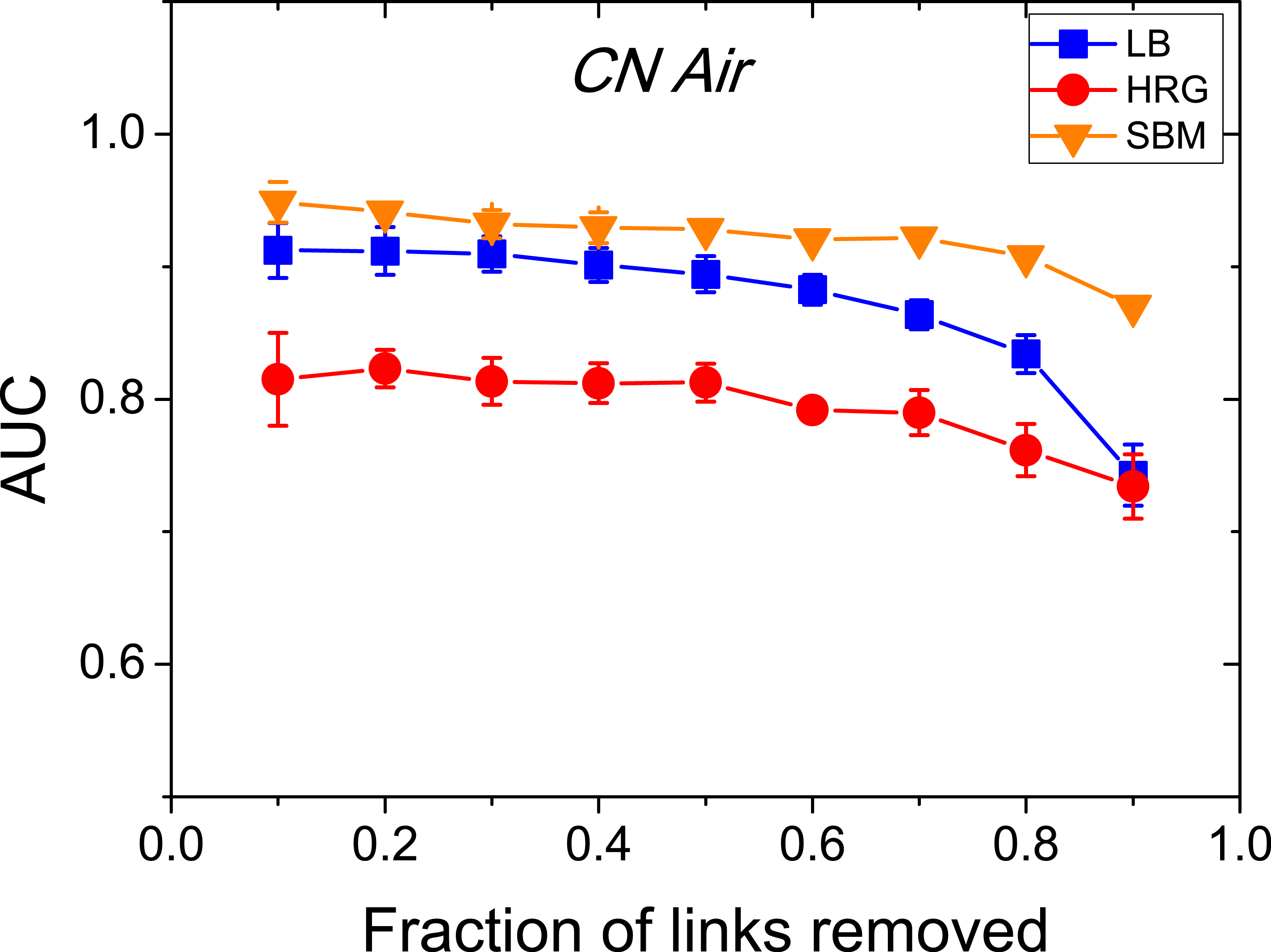}
\includegraphics[width=2.2in]{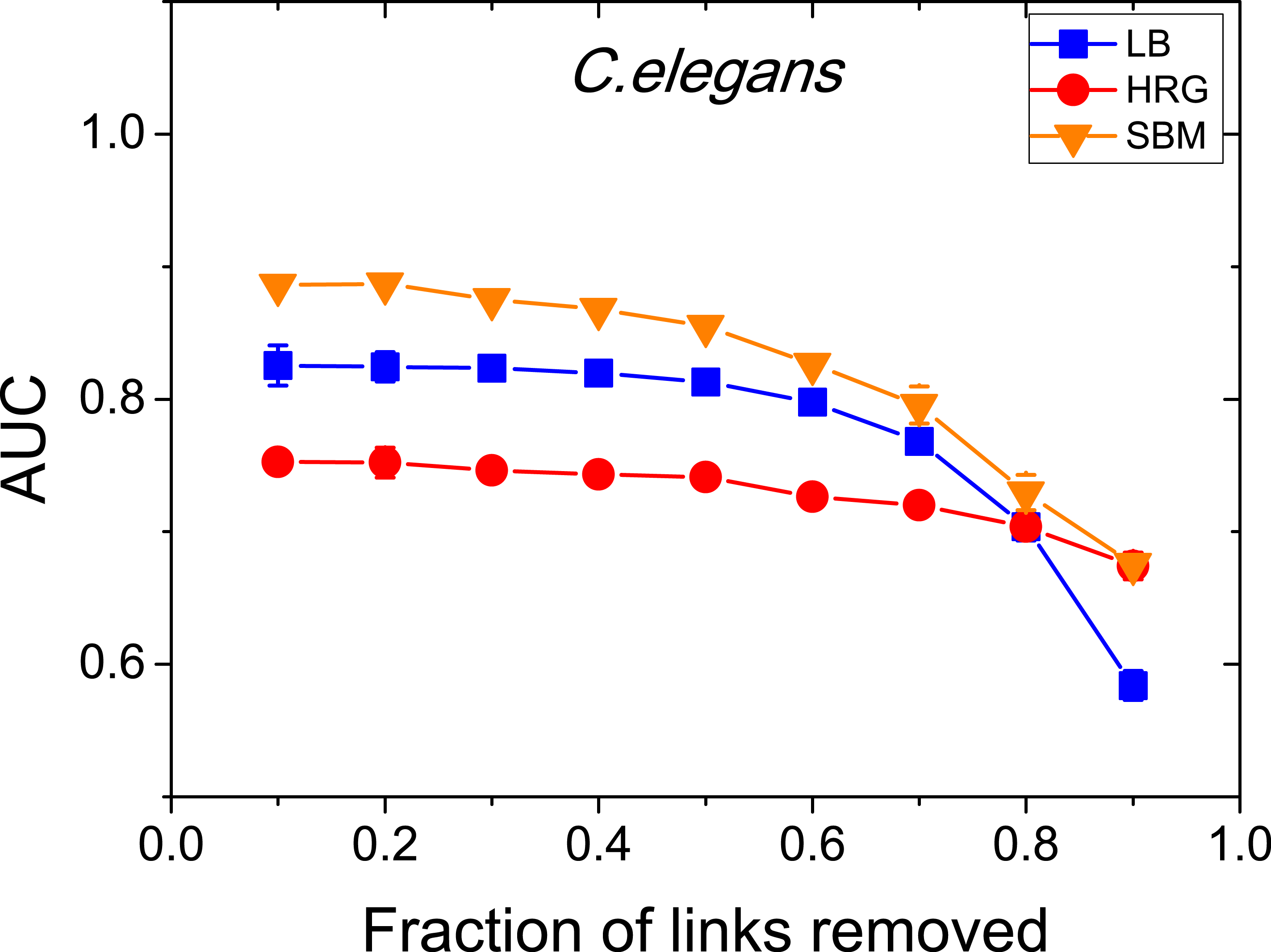}
\includegraphics[width=2.2in]{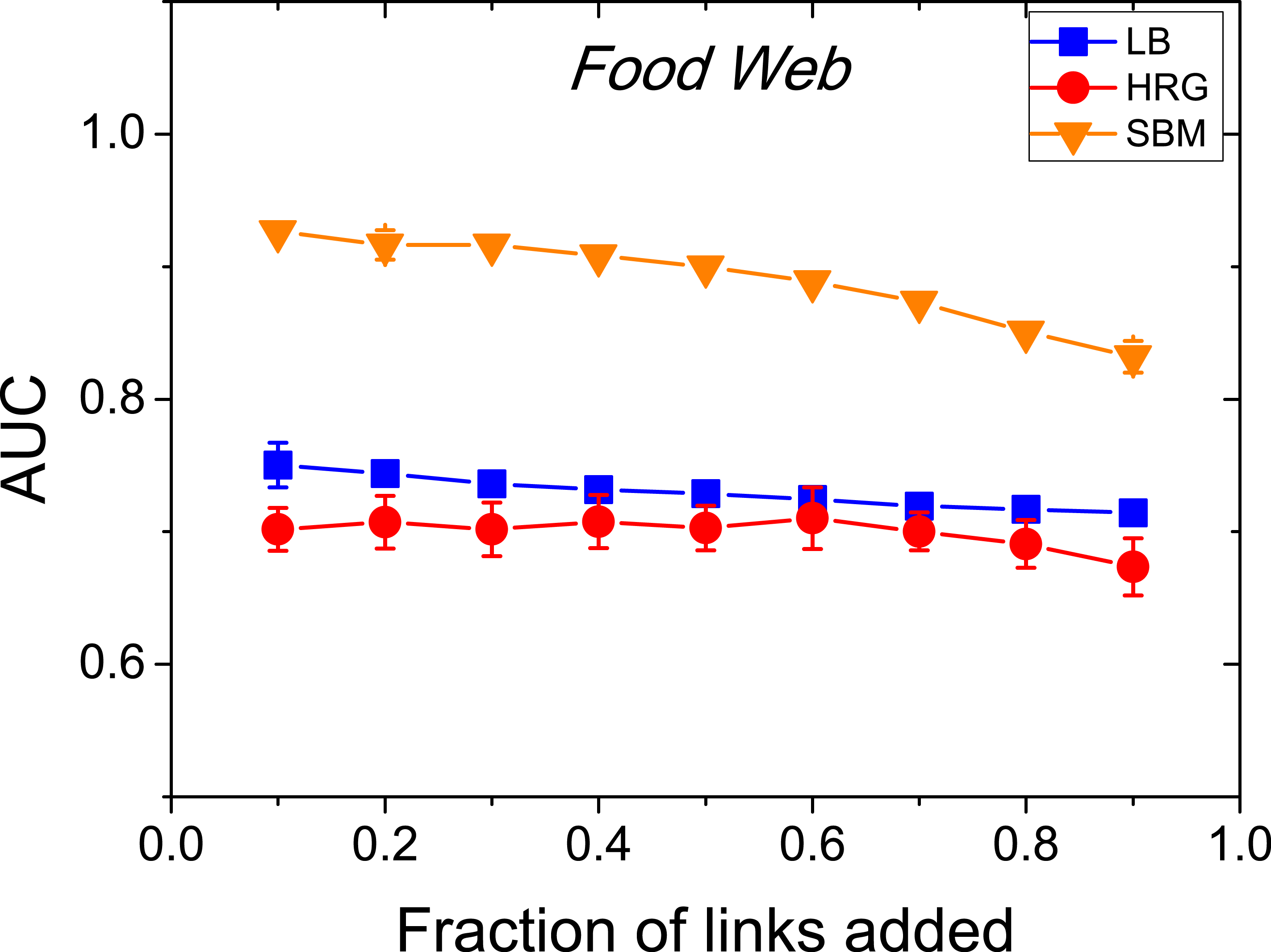}
\includegraphics[width=2.2in]{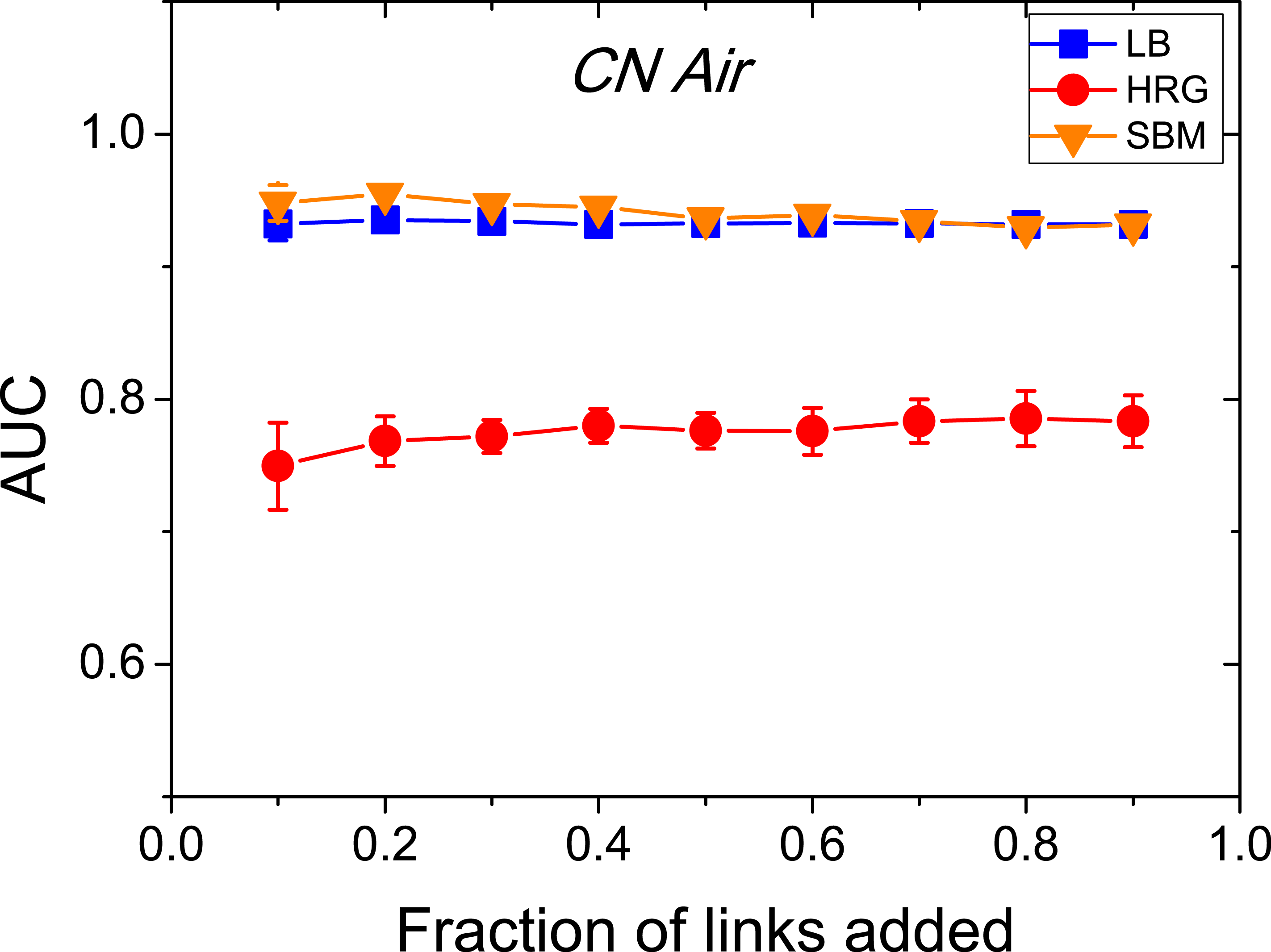}
\includegraphics[width=2.2in]{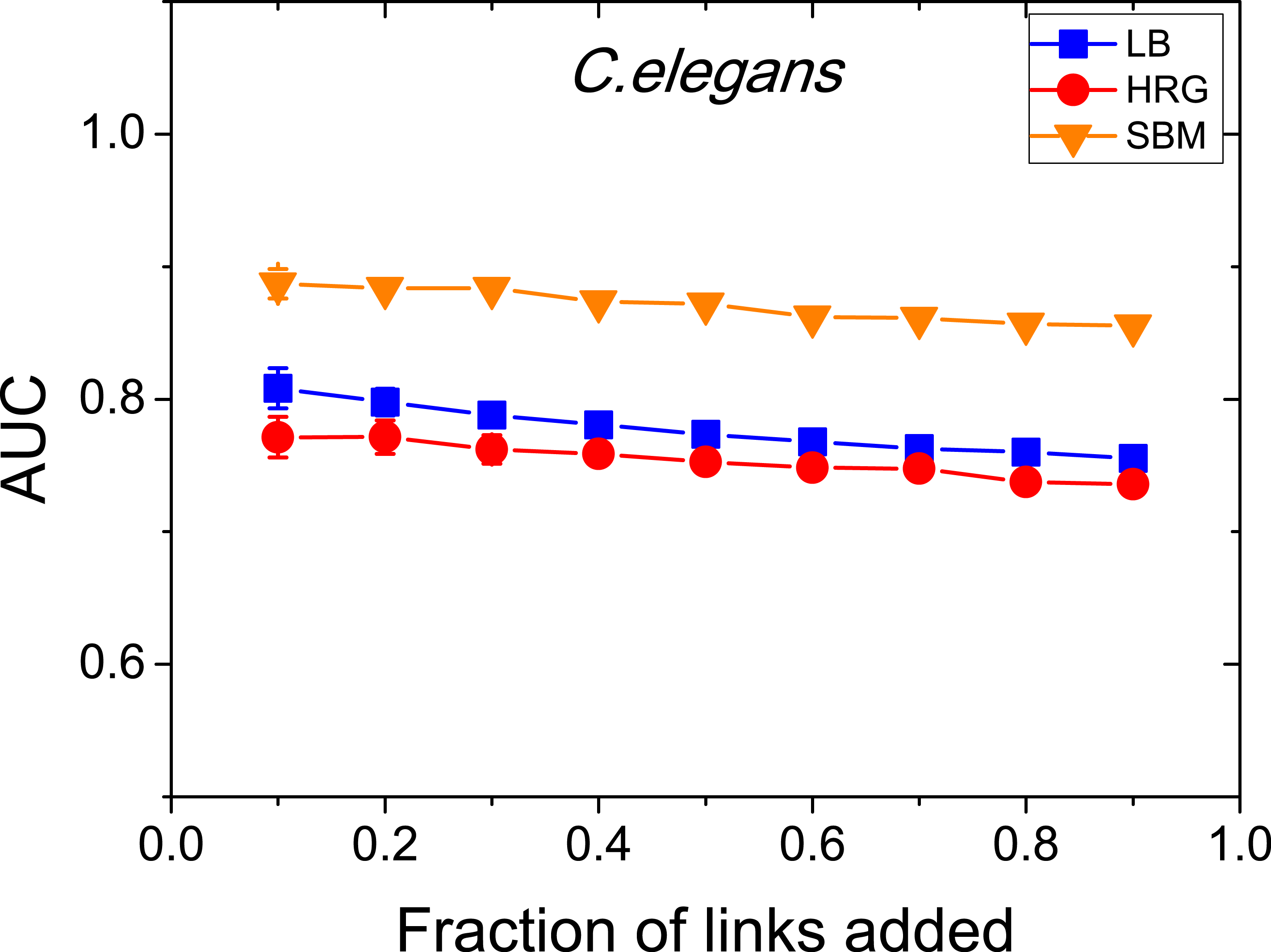}

\caption{\label{fig:figure4} AUC curves for missing link prediction (upper panel) and spurious link prediction (bottom panel). LB and two global predictors are compared by the experimental results performed on three networks. Each value of AUC is a result averaged over 100 tests and the error bar corresponds to the standard deviation.}
\end{figure*}

To futher evaluate the performance of LB index, we also implemented accuracy comparisons between our model and global-information-based link predictors like HRG, SBM on three relatively small networks including Food web, CN Air and C.elegans. The comparisons on larger-sized network are abandoned for a tremendous amount of calculating time required for HRG and SBM. To demonstrate this, the running time records of experiments on the three networks for all tested approaches are summarized in Table III. The results suggest that the LB index appears a similar time scale to other local methods while HRG and SBM are considerably time-consuming. According to Fig. \ref{fig:figure4}, the LB index overall preforms worse than SBM yet better than HRG on all three tested networks. it is beyond our expectation that LB has a comparable accuracy performance to global predictors since a local index commonly has a small chance to perform better than any global predictors in related literatures\cite{clauset2008hierarchical,guimera2009missing,liu2013correlations}.\\ 

\begin{table*}[htb]
\caption{\label{tab:table3} Comparisons of computational efficiency among seven predictors. Each value is the cumulative time in seconds for 100 experiments with 10\% fraction of links removed. The machine used for testing is a desktop with a processor of Intel (R)
Core (TM) i3 CPU 3320 @ 3.3 GHz and 16 Gigabytes memory.}
\centering
\begin{ruledtabular}
    \begin{tabular}{cccccccc}
    ~ & LB & CN &  AA & RA & PA & HRG & SBM \\ \hline
   Food web   & 1.70 & 1.2 & 1.19 & 1.19 & 1.36 & 8316.62 & 11841.43\\
   CN Air  & 3.72 & 2.25 & 2.31 & 2.32 & 2.18 & 13501.44 & 9958.76\\
   C.elegans  & 5.92 & 2.96 & 3.15 & 3.03 & 4.09 & 25554.94 & 87676.04\\
    \end{tabular}
\end{ruledtabular}
\end{table*}

\section{Analysis}
\begin{figure*}[htb]
\includegraphics[width=2.2in]{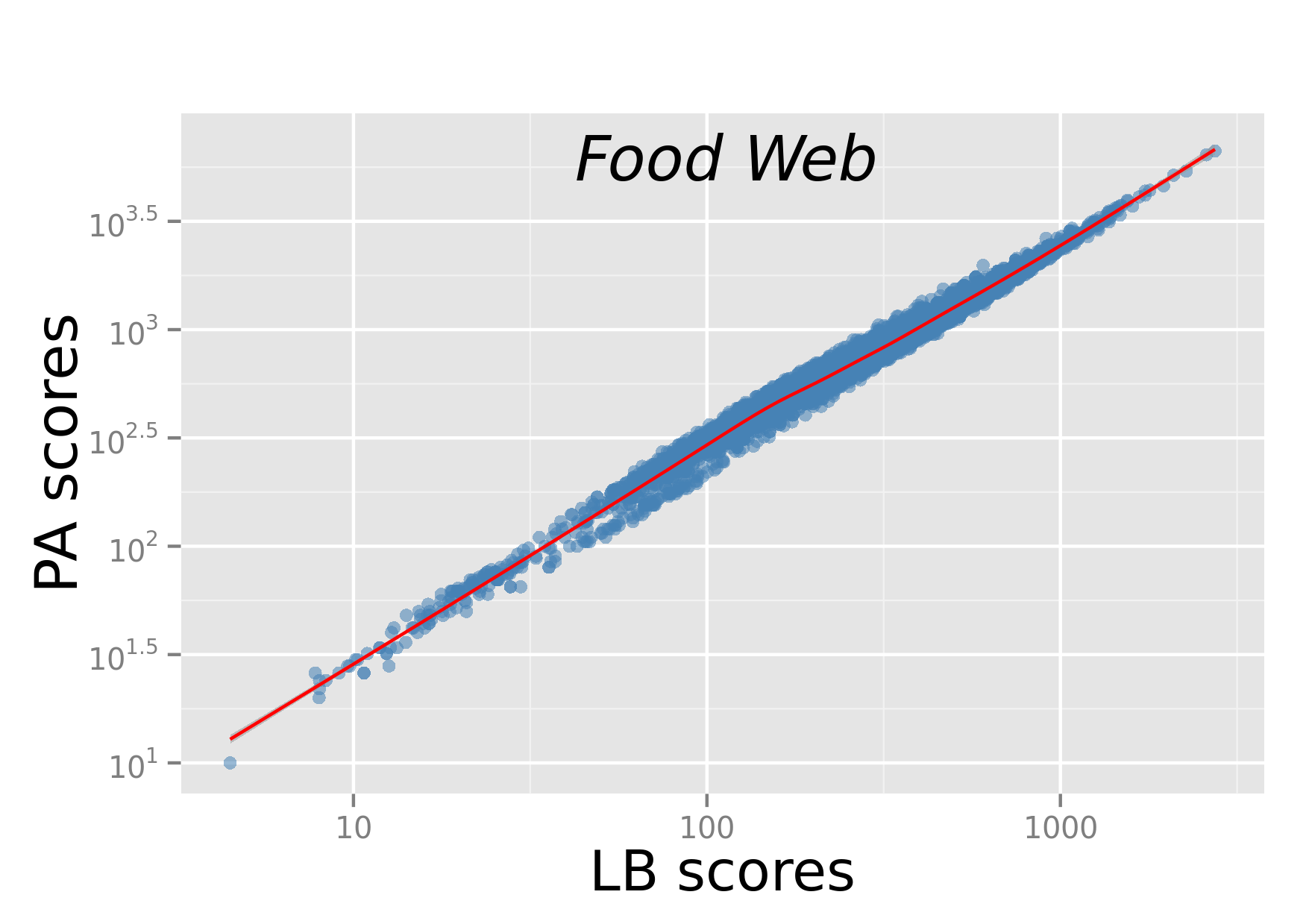}
\includegraphics[width=2.2in]{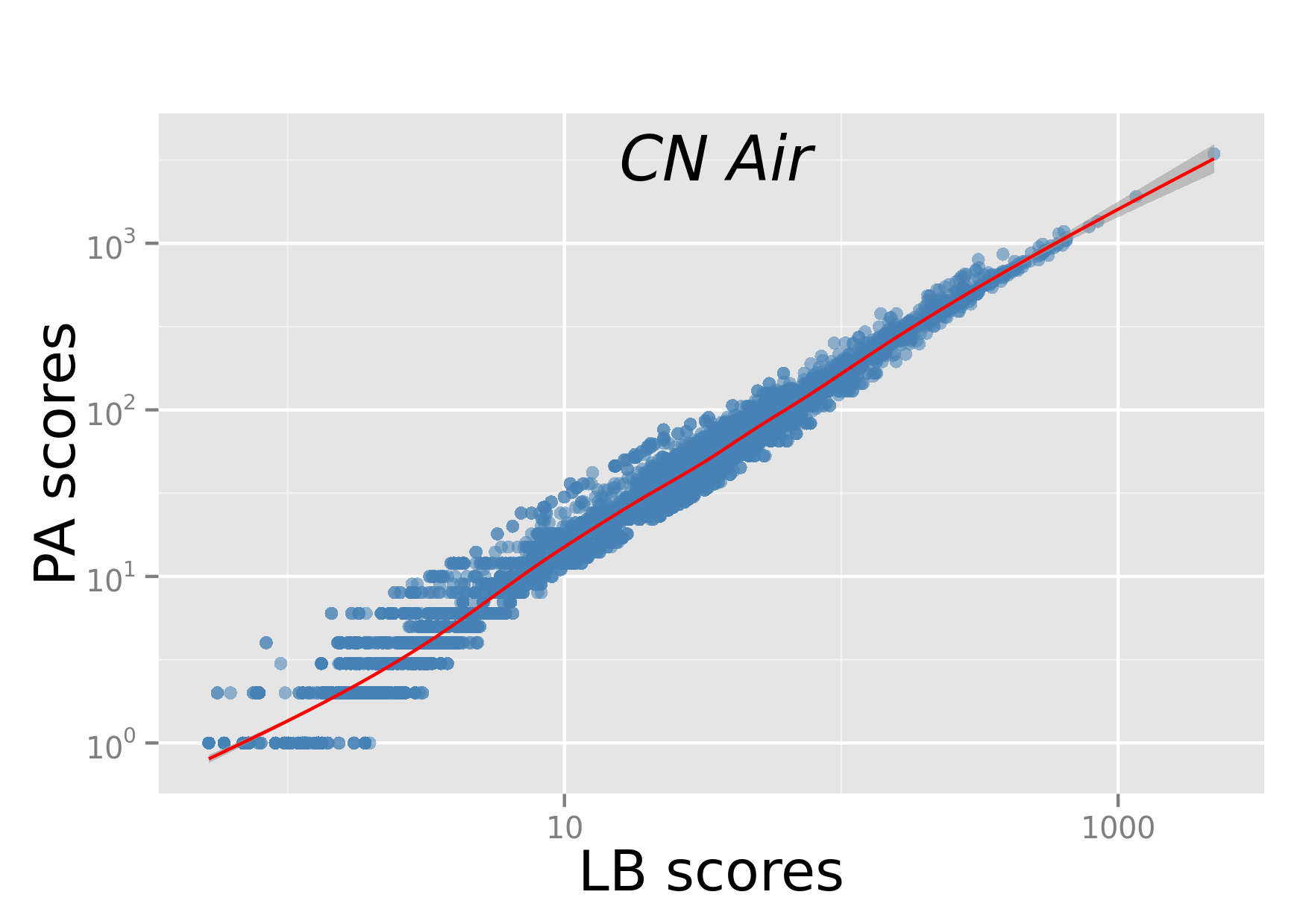}
\includegraphics[width=2.2in]{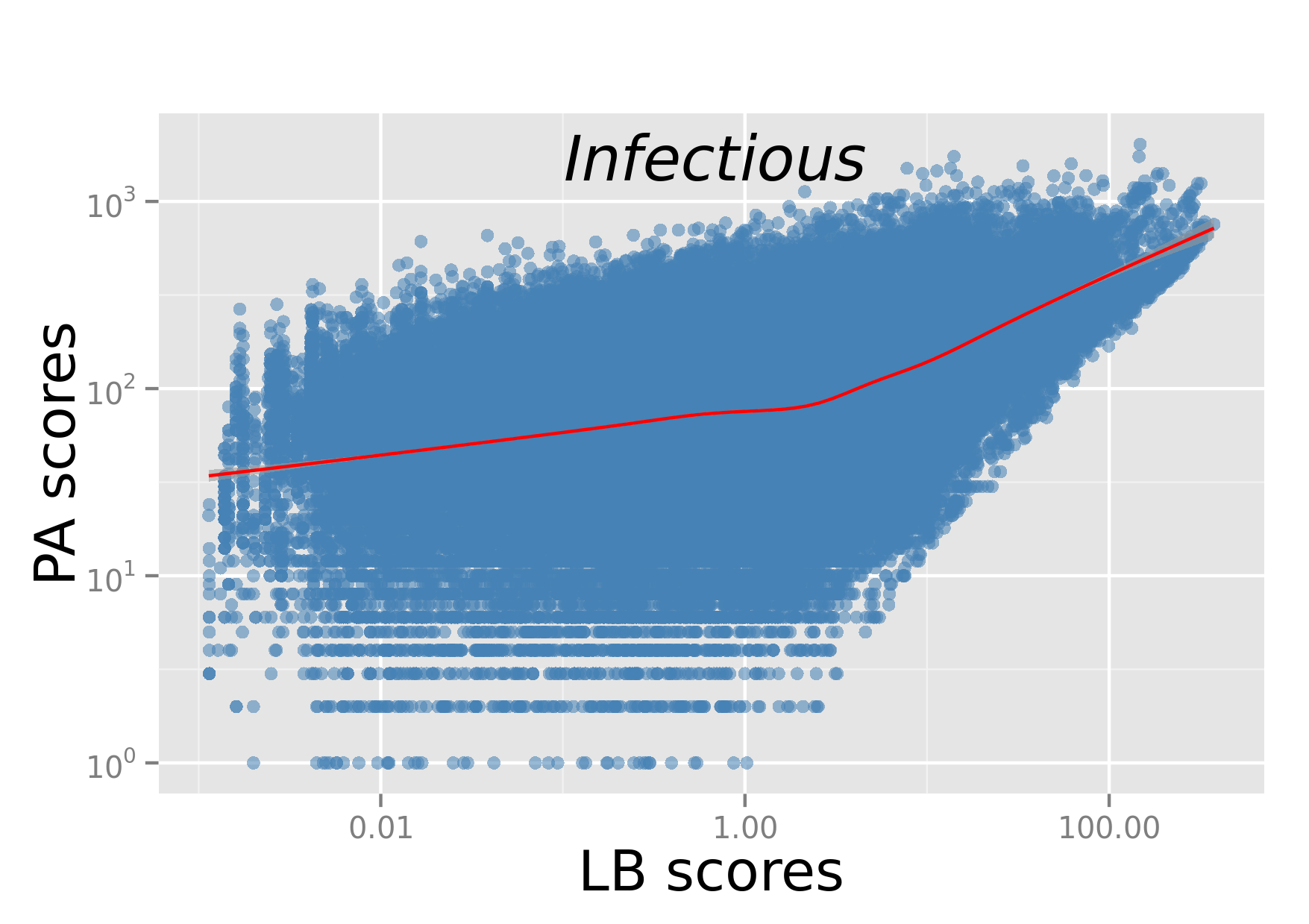}
\includegraphics[width=2.2in]{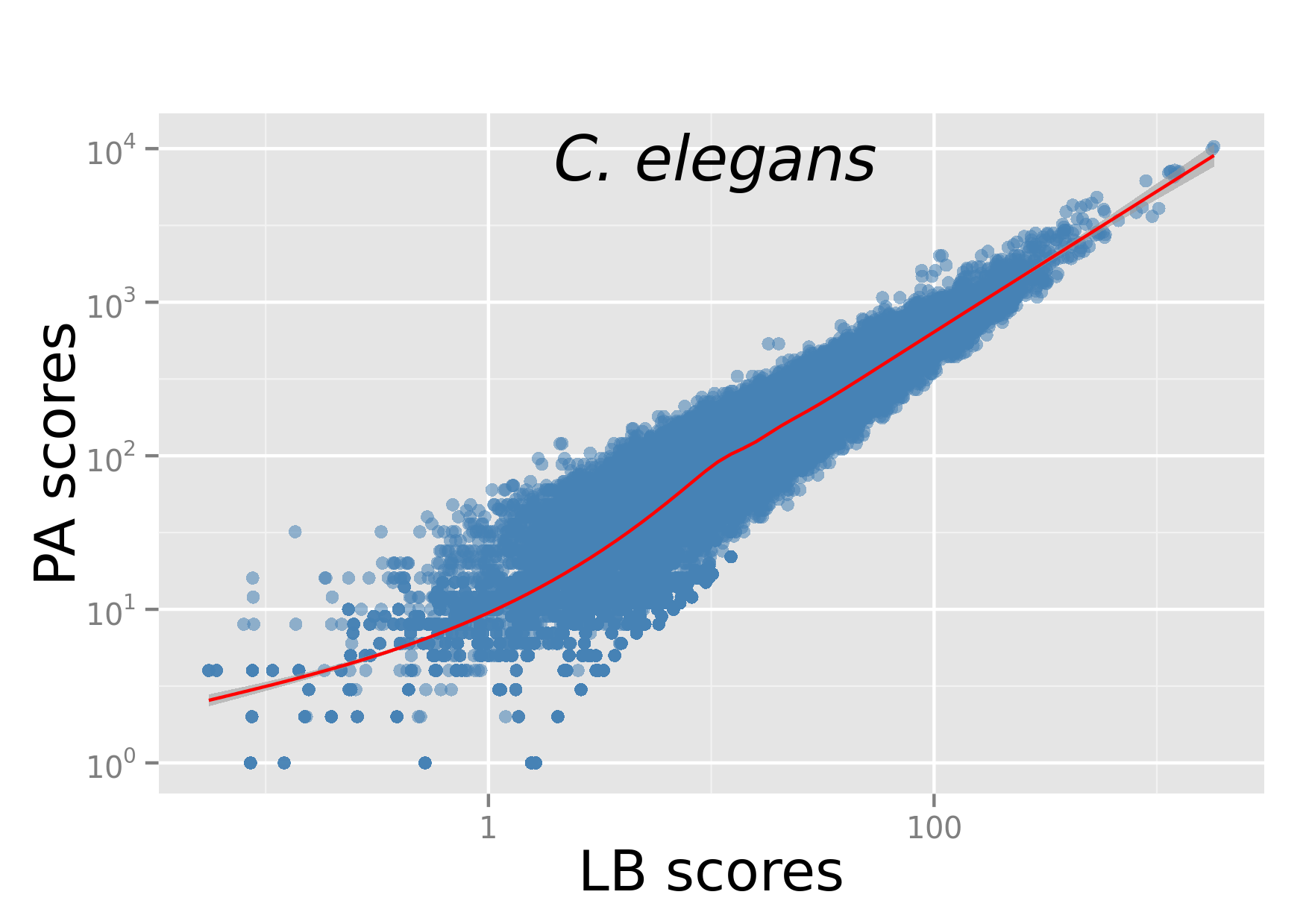}
\includegraphics[width=2.2in]{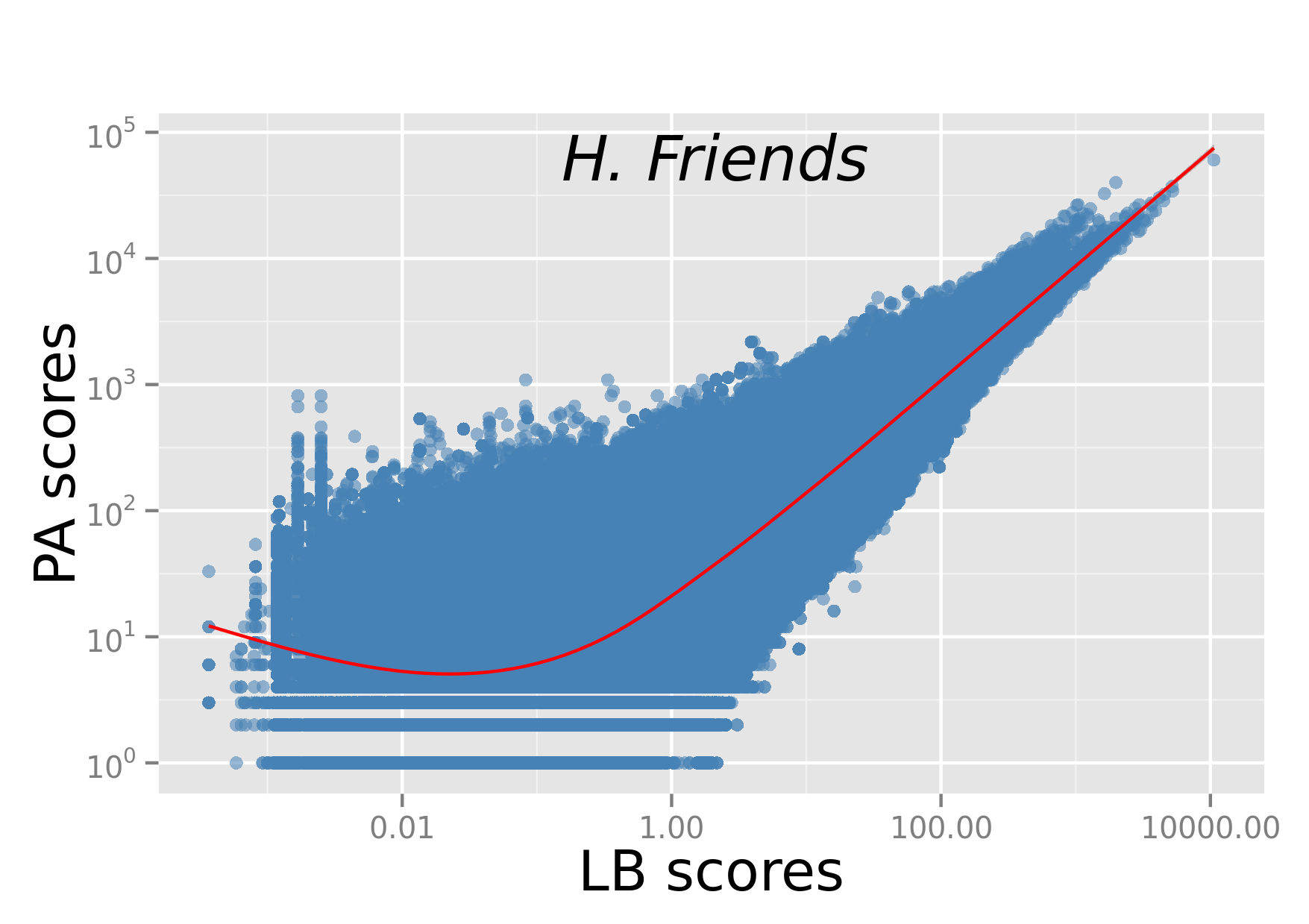}
\includegraphics[width=2.2in]{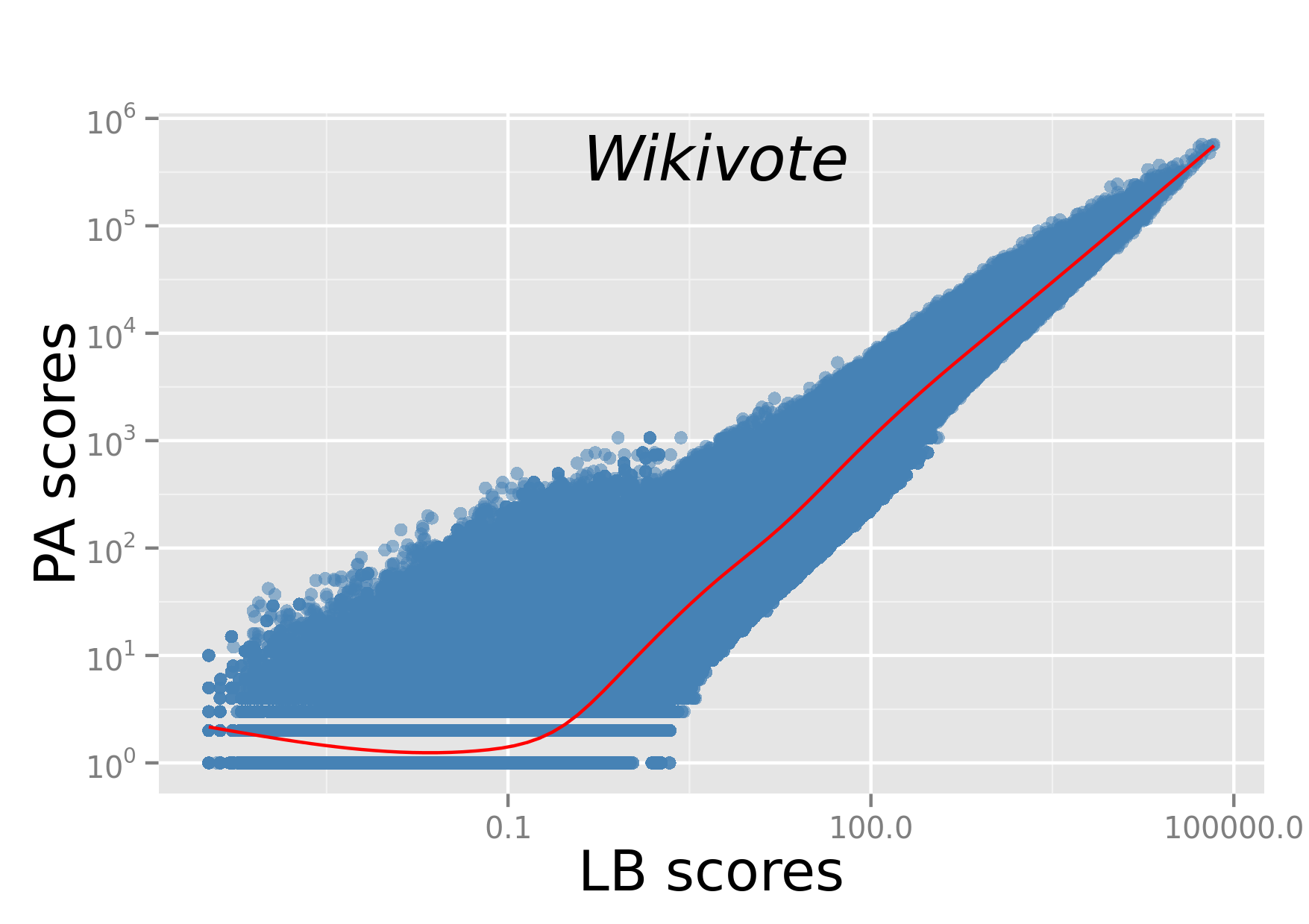}
\caption{\label{fig:figure5} Correlation distribution between PA scores and LB scores and the red lines are the trend line fitted by plot smoothing of Generalized Additive Model\cite{wood2000modelling} (The generalized additive model is a generalized linear model widely used in statistical analysis and applications). The shaded area indicates the level of confidence interval which is an estimated range to which the fitted curve belongs with a specified probability (0.95 adopted here). The distribution of PA scores and LB scores is plotted in double-logarithmic scale.}
\end{figure*}

\begin{figure*}[htb]
\includegraphics[width=2.2in]{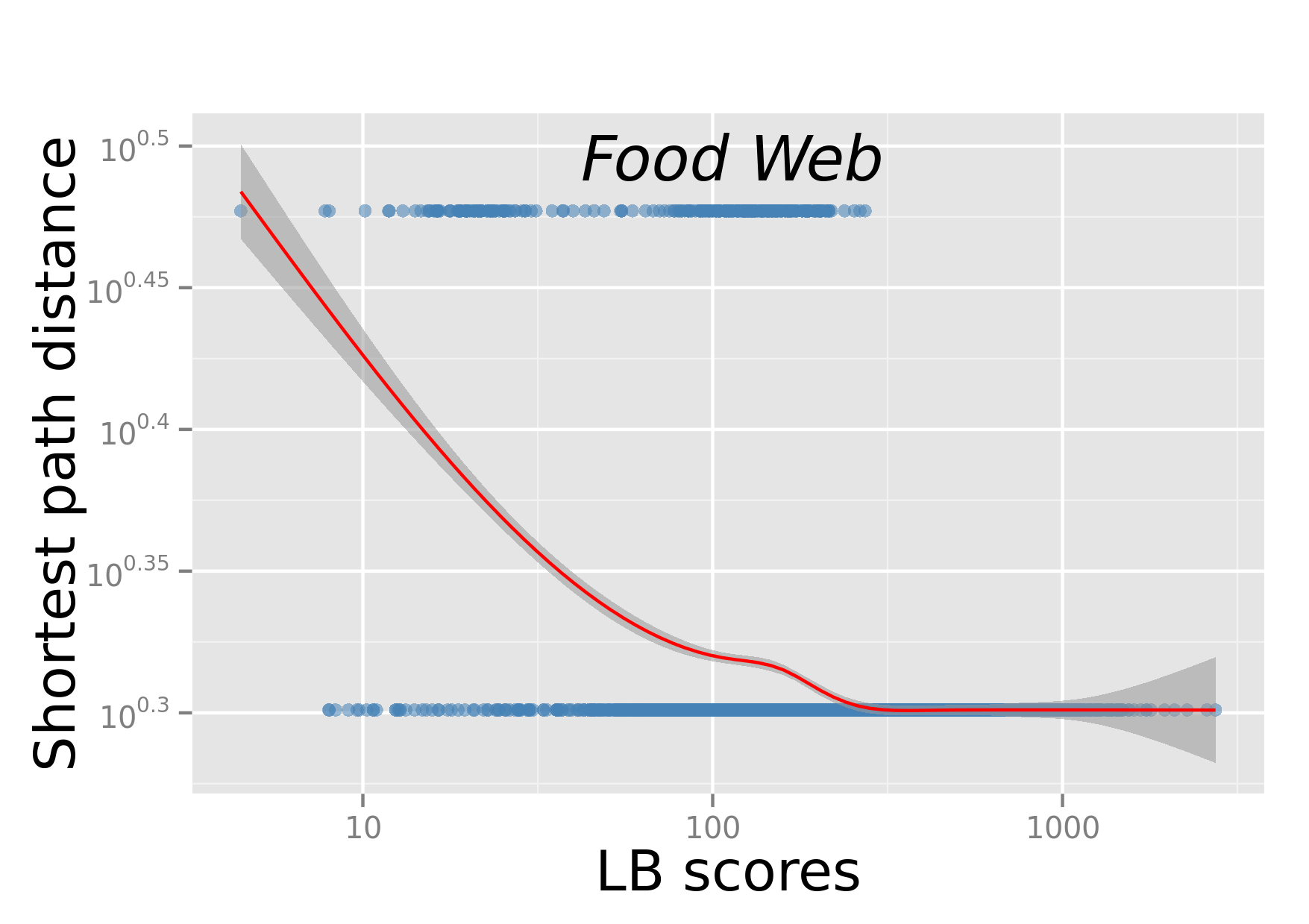}
\includegraphics[width=2.2in]{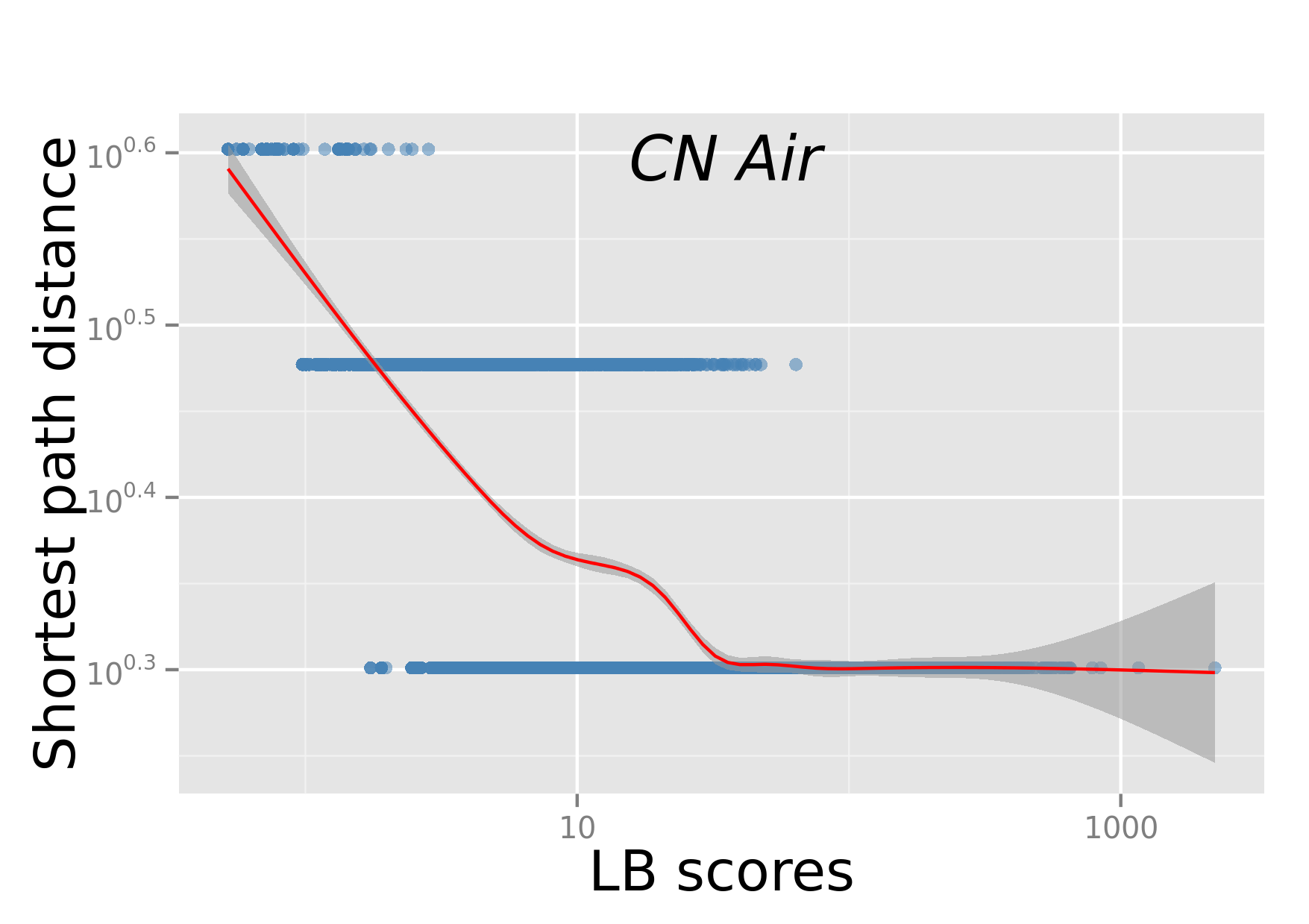}
\includegraphics[width=2.2in]{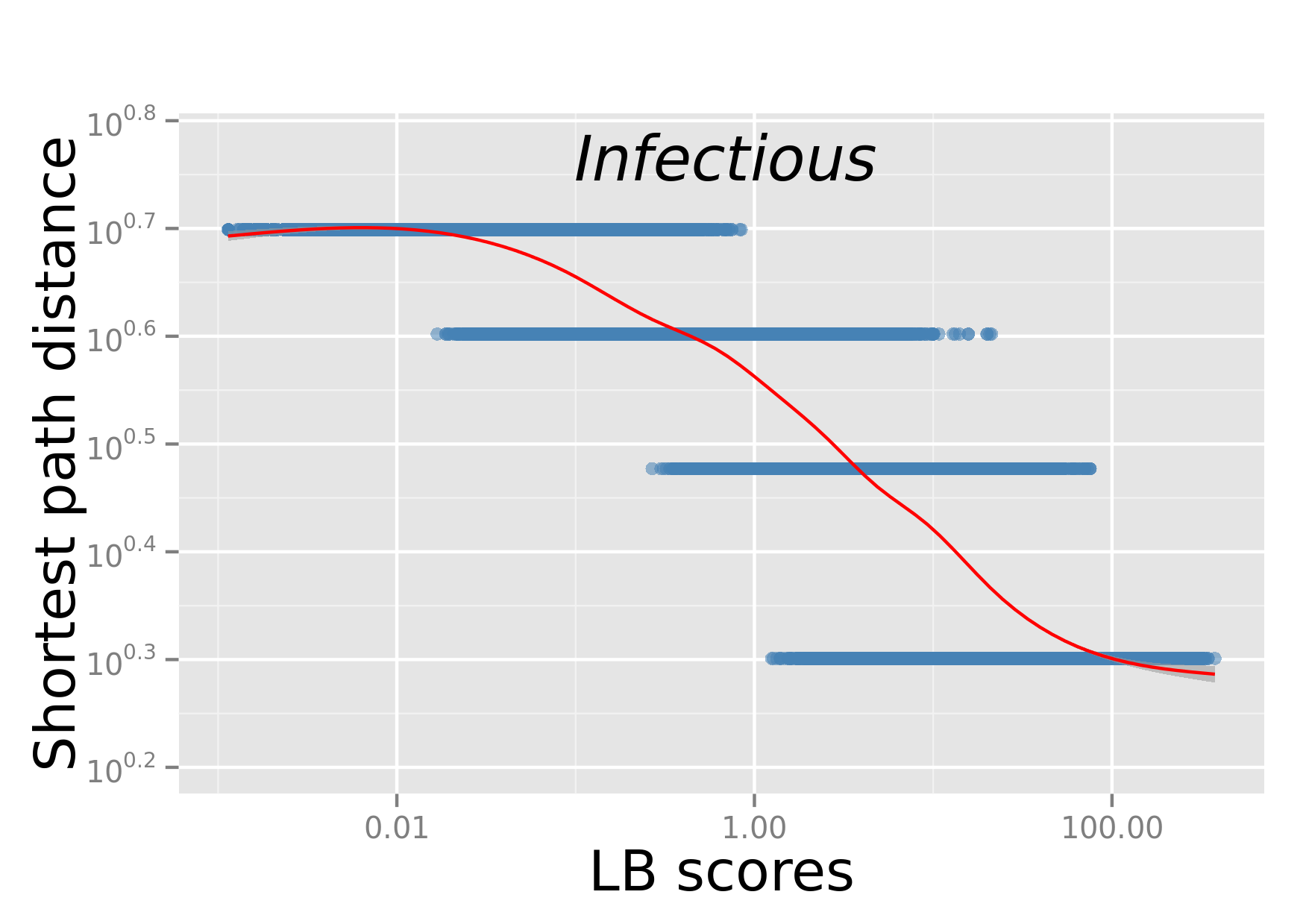}
\includegraphics[width=2.2in]{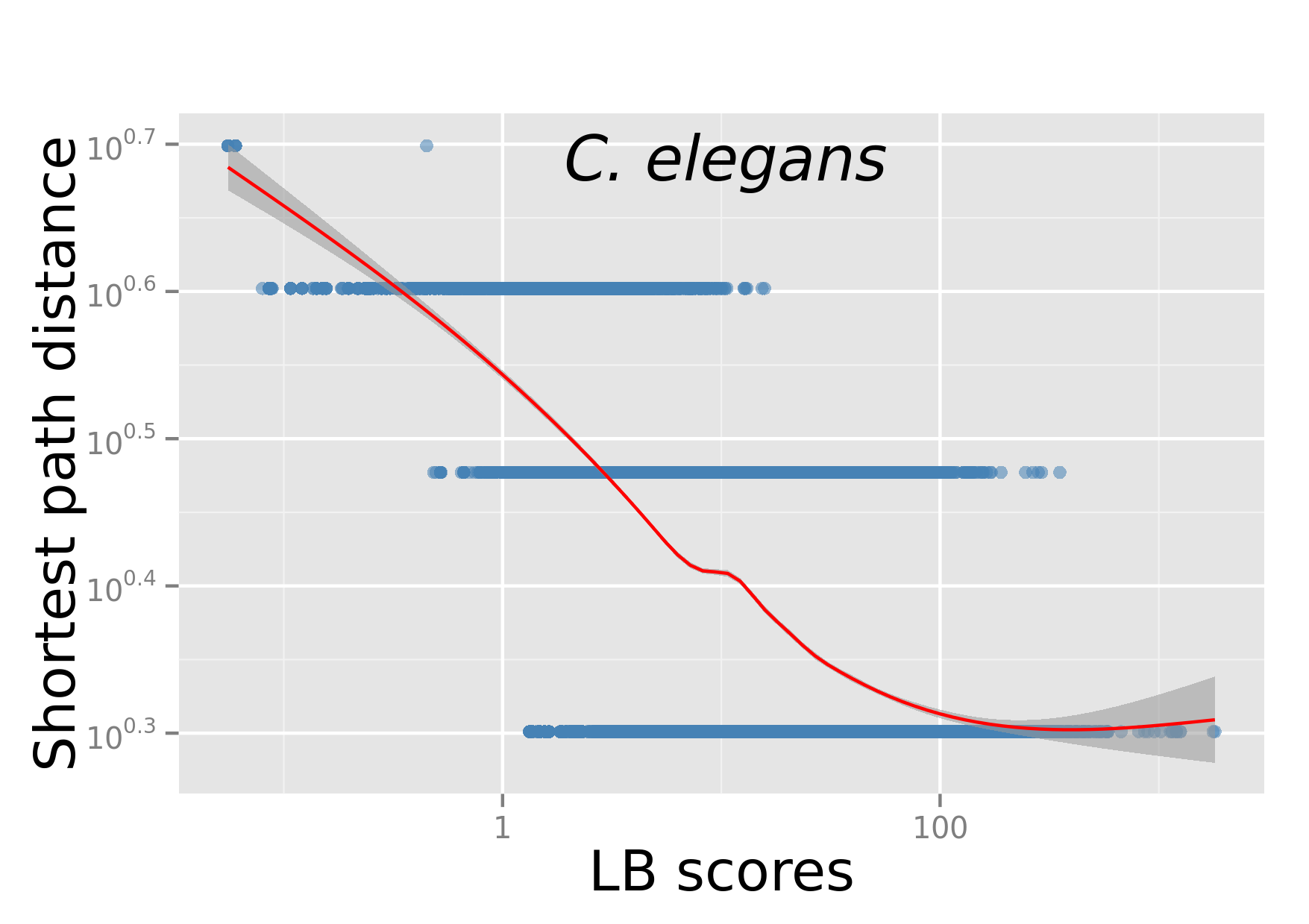}
\includegraphics[width=2.2in]{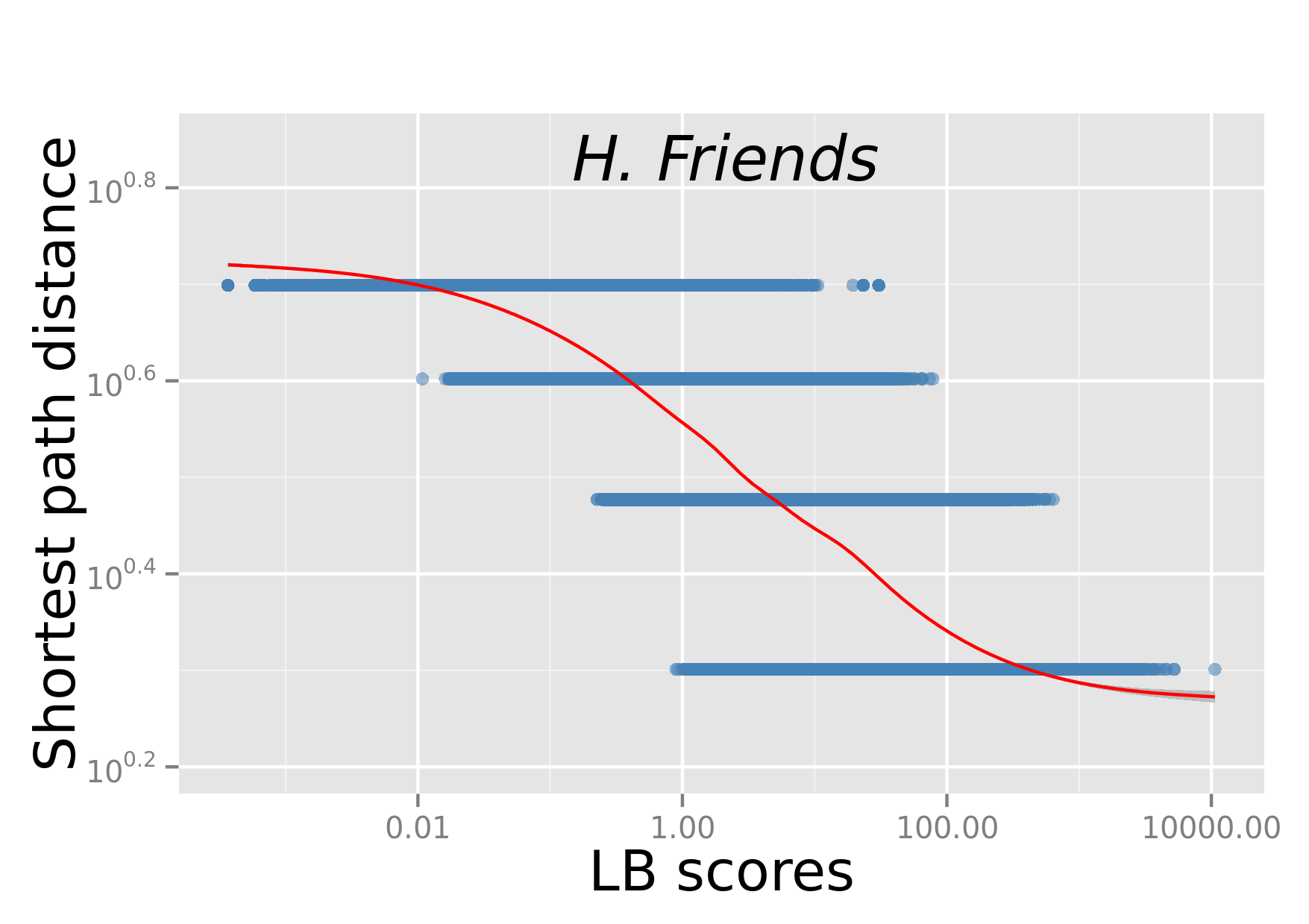}
\includegraphics[width=2.2in]{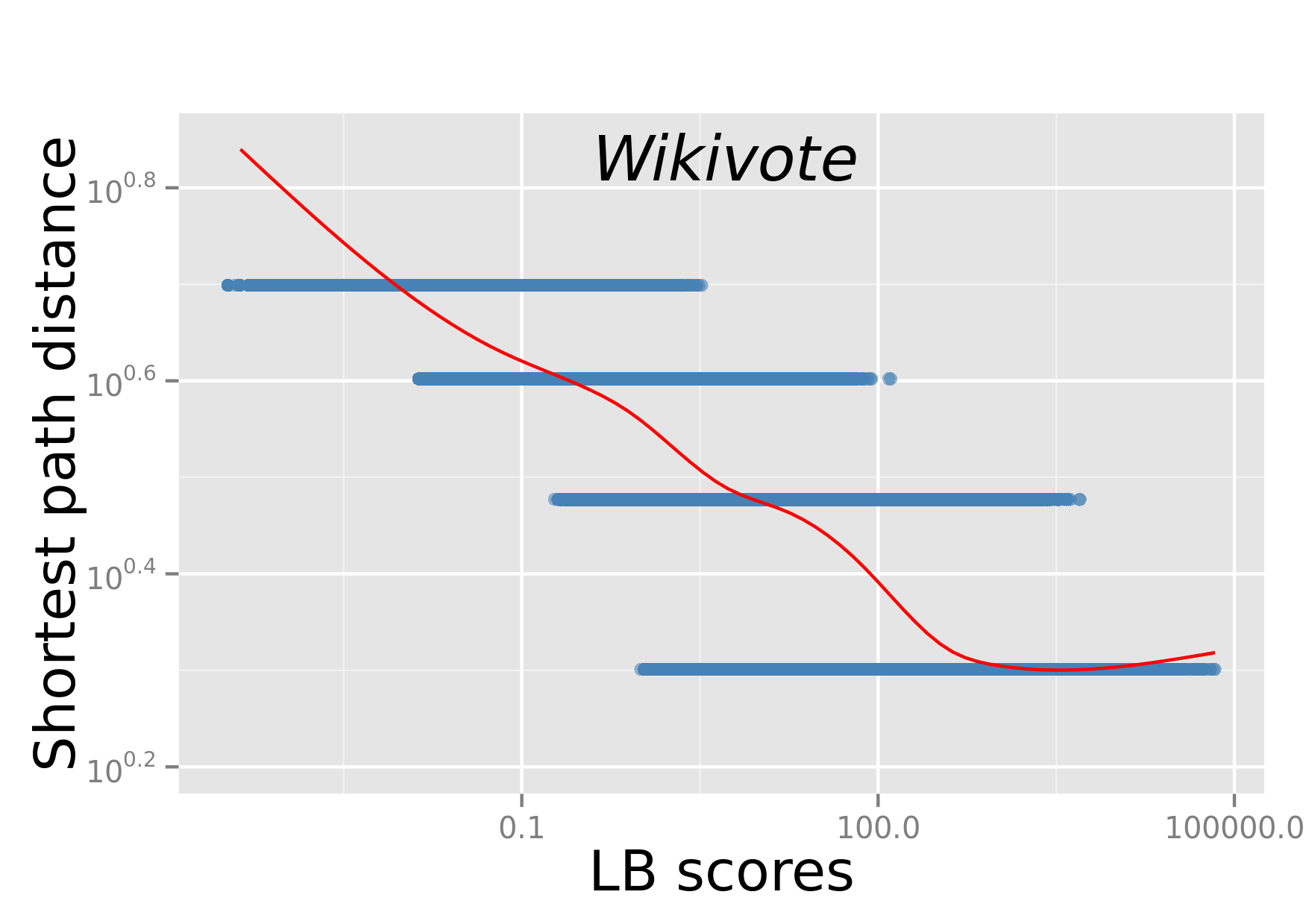}

\caption{\label{fig:figure6} Correlation distribution between Shortest path distances and LB scores for all node pairs and the red lines are the trend line fitted by plot smoothing of Generalized Additive Model. The shaded area indicates the level of confidence interval which is an estimated range to which the fitted curve belongs with a specified probability (0.95 adopted here). The distribution of shortest path distances and LB scores is plotted in double-logarithmic scale. }
\end{figure*}

Due to the remarkable performance of prediction shown by the LB index, we believe that the new index has captured some latent topology features which impact on the link formation in complex networks.  By analyzing the correlations between LB scores and those scores obtained by some other local predictors such as CN, AA, etc.,  to our surprise, we find that the similarity scores given by LB are positively correlated with those scores given by PA index shown in Fig. \ref{fig:figure5}. This indicates that the scores of LB index, to some extent, are analogous to those of PA index, and implies that the large-degree-favored link formation can be captured by the LB index as well. But according to the comparison results shown in Fig. \ref{fig:figure2} and Fig. \ref{fig:figure3},  LB index apparently performs better than the PA index on most tested networks. After further investigating the Fig  4, we notice that those pairs of nodes having identical PA scores usually have non-identical LB scores. This implies that LB index may have captured some other important link formation factors. By observing the topology difference among these pairs of nodes having identical PA scores, we find that the main difference among them is the shortest path distance. The correlations between shortest path distance and LB scores for all node pairs on the six networks are illustrated in Fig. \ref{fig:figure6}. The results indicate that, generally speaking, LB scores are negatively correlated with the distances of shortest path between pairs of nodes. In other words, pairs of nodes having longer shortest path distances would obtain lower LB scores. In fact, the short path represents a basic idea of path-based proximity predictors, such as Katz\cite{katz1953new} and Local path (LP)\cite{zhou2009predicting}, which means that the LB index would be similar to these predictors as well. Qualitatively speaking, a score given by the LB index demonstrates that a pair of nodes having larger degrees and shorter shortest path between them will be more similar to each other, thus, the two nodes are more likely to create a link.

According to our correlation analyses, we conclude that the link formation in real-world networks has two typical scenarios: i) The number of a node's neighbors commonly represents its activity. That is to say, a node is more active with more neighbors. Therefore, two active entities in a given network are very likely to form a link.  This can be summarized as large degree principle; ii) Long topology distance would hinder two nodes to form a link. The shorter the shortest path distance between two vertices is, the higher possibility a link has to be established between them. This can be summarized as short path principle. Of course, the two principles may play different roles in different networks. For example, in social networks, large degree nodes may play a more important role because popular nodes or active nodes may attract much attention from other nodes, which is easier to be captured by the PA index. However, in infectious networks, interactive infections would be the results of neighborhood influence within a short distance (commonly face to face). Therefore, in social networks, link formation procedure would be large degree first, whereas, in infectious networks, link formation procedure would be short distance first. This is able to explain why PA index performs pretty bad in infectious network shown in Fig. \ref{fig:figure2} and Fig. \ref{fig:figure3} and the correlation distribution between LB scores and PA scores in Infectious network is more scattered than other results in the rest networks shown in Fig. \ref{fig:figure5}. In most cases, link formation would be the consequence of joint influence by multiple mechanisms\cite{liu2013correlations,holme2002growing}. For example, besides the impact of large degree nodes, Leskovec et al's study also shows that most new links in social networks span very short distances, typically closing triangles\cite{leskovec2008microscopic}. Because of the two aspects of link formation simultaneously captured by the LB index in a balanced way, it has the capacity of performing better link prediction than other single-mechanism-driven methods such as traditional local indices.

% Figure 

\section{Conclusions}
In this article, we proposed a degree blocking model by using network's local topology information and carried out extensive experiments of link prediction on various real-world networks including an ecological network, two social networks, a transportation network, a biological network and an epidemic contact network. Experimental results validate that, in most real-world networks, the new index has better accuracies than other local-similarity-based indices. According to our analyses, the LB index is essentially a balanced hybrid version of PA index and short-path-based index, which can capture both large degree principle and short path principle. In our opinion, to design a new link predictor by simply combining the two types of indices together would not adapt to a wide range of networks because the two aspects usually play different roles in different networks and the binding or coupling pattern would be complex (commonly nonlinear). However, a significant merit of the LB index is that it is able to fit the two aspects, without parameter tuning, to various kinds of networks properly in terms of our experiments. To the best of our knowledge, the LB index is a firstly proposed multi-mechanism driven link predictor which has presented some new evidences to support a widely accepted viewpoint that the process of link formation in complex networks would  be a result of the joint influence of several mechanisms. Our work has provided new insights for researchers in developing some simple yet effective link prediction models in the future.

\begin{acknowledgments}
We wish to thank the funds of National Natural Science Foundation of China under grant Nos. 60903073, 61103109, the research funds for central universities under grant No. ZYGX2012J085 and Special Project of Sichuan Youth Science and Technology Innovation Research Team (2013TD0006).
\end{acknowledgments}

% The \nocite command causes all entries in a bibliography to be printed out
% whether or not they are actually referenced in the text. This is appropriate
% for the sample file to show the different styles of references, but authors
% most likely will not want to use it.
\nocite{*}

\bibliography{local_degree}% Produces the bibliography via BibTeX.

\end{document}